\documentclass[12pt]{article}
\pdfoutput=1

\usepackage{epsfig}
\usepackage{graphicx}
\usepackage{color}
\usepackage[dvipsnames]{xcolor}
\usepackage{caption}
\usepackage{subcaption}
\usepackage[numbers,compress]{natbib}
\usepackage[T1]{fontenc} 
\usepackage{amsmath,amssymb,mathrsfs}
\usepackage[colorlinks=true
,urlcolor=blue
,anchorcolor=bleu
,citecolor=blue
,filecolor=blue
,linkcolor=blue
,menucolor=blue
,pagecolor=blue
,linktocpage=true
,pdfproducer=medialab
,pdfa=true
]{hyperref}

\definecolor{mathblue}{rgb}{0.368417, 0.506779, 0.709798}

\def\be{\begin{eqnarray}}
\def\ee{\end{eqnarray}}
\def\nn{\nonumber\\}

\def\ct{\cite}
\def\la{\label}
\def\eq#1{(\ref{#1})}

\def\a{\alpha}
\def\b{\beta}

\def\d{\delta}
\def\D{\Delta}
\def\e{\epsilon}

\def\L{\Lambda}
\def\m{\mu}
\def\n{\nu}

\def\r{\rho}

\def\pa{\partial}
\def\fr{\frac}

\def\lc{\left\{}
\def\ls{\left(}
\def\lp{\left.}

\def\rc{\right\}}
\def\rs{\right)}

\begin{document} 
\flushbottom
\begin{titlepage}

\hfill\parbox{5cm} { }

\vspace{25mm}

\begin{center}
{\Large \bf Holography Transformer}

\vskip 1. cm

Chanyong Park$^a$\footnote{e-mail : cyong21@gist.ac.kr},  
Sejin Kim $^b$\footnote{e-mail : sejin817@kookmin.ac.kr} and Jung Hun Lee$^b$\footnote{e-mail : junghun.lee@kookmin.ac.kr}

\vskip 0.5cm

{\it  $^a\,$Department of Physics and Photon Science, Gwangju Institute of Science and Technology, Gwangju 61005, Korea}\\
{\it  $^b\,$College of General Education, Kookmin University, Seoul, 02707, Korea}\\
\end{center}

\thispagestyle{empty}

\vskip1cm


\centerline{\bf ABSTRACT} \vskip 4mm

We have constructed a generative artificial intelligence model predicting the gravity solutions when the holographic entanglement entropy of the dual quantum field theory is provided as input. The model we utilize is based on the transformer algorithm commonly used in natural language tasks such as text generation, summarization, and translation. The transformer model can understand the implicit relation between input and output sequences by training data. For training, we generate many data sets composed of holographic entanglement entropy and corresponding metric solutions. After training these data, the transformer model predicts the dual geometry from arbitrary test sets of entanglement entropy data. The reconstruction of the dual gravity allows us to get more information on the thermodynamic quantities of thermal systems, which cannot be read directly from entanglement entropy data. In this work, we construct the dual geometry by applying the transformer model. After that, we derive thermodynamic quantities, like temperature and densities of thermal systems, from the entanglement entropy.  \\

\vspace{2cm}


\end{titlepage}


\section{Introduction}

From ancient times to the present, physics has long aimed to try to understand and explain our world through mathematical description. In the past, Newton's laws of motion were used to describe the movements of objects, enabling precise calculations of their positions at specific times and the analysis of forces acting upon them. However, as modern science progressed, Newton's world view has reached its limit as a good tool, in the process of overcoming this circumstance, new theories like quantum mechanics and the theory of relativity have been developed.

Quantum gravity is a new theory that provides a unified viewpoint to look at gravity and the quantum theory. Especially, the holographic theory so-called anti-de Sitter space/conformal field theory (AdS/CFT) correspondence has been accepted as the most probable candidate for describing them together. The core idea of the AdS/CFT correspondence is that the information of quantum states can be encoded in one-higher dimensional spacetime \cite{Maldacena:1997re, Gubser:1998bc, Witten:1998aa, Witten:1998ab, Aharony:1999ti,  deHaro:2000vlm, Benna:2008zy, Bergman:2018hin, Ryu:2006ef, Swingle:2009bg, Mojtahedi_2022}, where the additional holographic coordinate plays the role of energy scale from the viewpoint of the holographic renormalization group (RG) flow \cite{Park:2022aa, Park:2022fqy, Park:2020aa, Park:2018aa}. Based on the correspondence, various areas of physics have been extensively studied  \cite{Erlich:2005qh, Karch:2006pv, Karch:2002sh, Casini:2011kv, Faulkner:2013ana, Myers:2010tj}.

One of the important progress has been achieved in \cite{Ryu:2006ef}. It is well known that the computation of the entanglement entropy is a hard task on the field theory side. However, the holographic method proposed in \cite{Ryu:2006ef} provided an easier way to calculate the entanglement entropy and opened a new area to be investigated \cite{Faulkner:2013ana, Myers:2010tj, Maldacena:2013xja, Lewkowycz:2013nqa, Nishioka:2009un, Penington:2019npb}.

In holography studies, the integration of artificial intelligence (AI) provides researchers with a powerful tool to process large data sets and extract insights that may be challenging or impossible to obtain through traditional methods. Recently, AI has been employed to study holography theory, particularly in deep learning techniques applied to analyze strongly-coupled systems \cite{Park:2022fqy, You_2018, Hashimoto:2018wm, ShibaFunai:2018aaw, Akutagawa:2020aa, Chen:2021giw, Lam:2021ugb, Li:2023aa}. Deep neural networks have been used to predict and understand the behavior of condensed matter systems, leading to discoveries of new phase transitions. AI has also been utilized to analyze data from QCD simulations, providing insights into the behavior of strongly coupled systems in this realm. Additionally, AI algorithms have been employed to analyze data from simulations involving the AdS/CFT correspondence, advancing our understanding of quantum gravity.
    
This paper focuses on the application of AI methods to holographic correspondence to reconstruct the dual gravity theory, especially from data on quantum entanglement phenomena. Our main purpose is to reconstruct theories or equations of motion from experimental or observational data. The equations of motion for quantum theory and gravity theory are composed of nonlinear partial differential equations, making it impossible to find theories that reproduce experimental results using methods other than trial and error. Especially for quantum theory, even if we know the equations accurately, there are only a few cases where we know the exact solutions.

To reconstruct holographic theories from given experimental results, we need to accurately analyze and understand the correlations between data points. We will utilize the transformer algorithm, a powerful artificial intelligence tool that can comprehend important information from data and generate new insights. 

In this paper, we will model a transformer to predict a geometric solution of three-dimensional gravity theory, while satisfying boundary conditions and experimental data. This model will incorporate transformer encoders, which will determine the relationship between the sequence points of the holographic entanglement entropy data and predict the gravity solution satisfying it in our metric ansatz.

We have organized our paper as follows. 
In section 2, we review the holographic method of calculating the entanglement entropy and explain the basic concept of transformer AI, and how to train our model to extract holographic data from the entanglement entropy. Section 3 is devoted to the application of our transformer model to three-dimensional black holes with $p$-brane gases. Finally, we close this work with some concluding remarks in section 4.


\section{Underlying principles for reconstructing dual geometries}

New macroscopic laws frequently occur in low-energy physics like condensed matter and nuclear physics, which cannot be explained by perturbative fundamental theories. To understand such new macroscopic phenomena, we must know a nonperturbative RG flow involving all quantum effects. However, it is not easy to figure out such a nonperturbative RG flow in traditional QFTs. In this situation, the holographic principle can shed light on this issue. According to the AdS/CFT correspondence, $d$-dimensional nonperturbative QFTs can be described by $(d+1)$-dimensional classical gravity theories. Moreover, thermal QFTs map to black hole geometries. In this case, the gravity theory can play the role of an effective theory. This implies that once the dual gravity is specified, it provides more information about the nonperturbative features of the dual QFT. For instance, when the entanglement entropy data is given, we can understand other physical properties by reconstructing its dual gravity theory. In the present work, we will discuss how one can determine thermodynamic properties from entanglement entropy by applying holography and machine learning techniques.

\subsection{Holography}

Before discussing the reconstruction of the dual geometry from entanglement entropy data, we first summarize the underlying theories, the holography and the transformer method. Let us consider a medium composed of several matters at finite temperature. A thermal system in the holographic setup is mapped to a black hole geometry. In this case, matters are characterized by black hole hairs. Therefore, a medium consisting of several matter is the dual of a black hole having multiple hairs. For example, charged black holes and $p$-brane gas geometries correspond to such multiple-hair black hole geometries. If we consider a bulk gauge field and $p$-branes extending to the radial direction of a background AdS space, a $p$-brane gas geometry appears as a geometric solution of the following gravity action
\be
S = \fr{1}{16 \pi G} \int d^{d+1} x \sqrt{-g} \ls {\cal R} - 2  \L   - F_{\m\n} F^{\m\n}  \rs +  \sum_{p=1}^{d-1} T_p  N_p \int d^{p+1} \xi \sqrt{- h} \ h^{\a \b} \pa_\a x^\m \pa_\b x^\n g_{\m\n} ,
\ee 
where $T_p$ and $N_p$ indicate the tension and number of $p$-branes, respectively. $F_{\m\n}$ is the field strength of a bulk gauge field $A_\m$. 

Assuming that $A_\m$ relies only on the radial coordinate and taking the following metric ansatz
\be
ds^2 = \fr{R^2}{z^2} \ls - f (z) dt^2 + \fr{1}{f(z)} dz^2 + \d_{ij} \, dx^i dx^j  \rs ,
\ee 
where $i,j = 1, \cdots, d-1$, the Euler equation of $A_\m$ allows the following solution
\be
A_\m (z) = \m + Q z^{d-1} .
\ee
Considering the gravitational backreaction of the bulk gauge field and $p$-branes, the Einstein equation determines the blackening factor to be
\be
f (z) = 1 - \sum_{p=1}^{d-1} \r_p \, z^{d-p} - M z^d + Q^2 z^{2(d-1)} ,
\ee
where $\r_p$ is associated with the density of $p$-branes. Here, $M$ and $Q$ correspond to the mass and charge of a black hole, respectively. In general, a black hole geometry satisfies the thermodynamics laws.
In the holographic setup, these thermodynamic properties describe the thermalization of the dual QFT. A multiple-hair black hole usually has many horizons satisfying $f(z_h) =0$. Denoting an outer horizon as an event horizon $z_h$, the black hole mass can be represented as a function of the other parameters, $\lc \r_p, Q, z_h \rc$. Then, the blackening factor in terms of these parameters is rewritten as
\be
f (z) = 1 - \sum_{p=1}^{d-1} \r_p \, \ls z^{d-p}  -  z_h^{-p} z^d \rs - z_h^{-d} \, z^d + Q^2 \ls z^{2(d-1)}  -  z_h^{d-2} \, z^d\rs .
\ee

For black hole geometries, the regularity at the event horizon defines the Hawking temperature as
\be
T_H = - \fr{ f'(z_h)}{4 \pi} .
\ee
Representing the event horizon as a function of temperature, $z_h ( \r_p, Q,T_H)$, the independent parameters $\lc \r_p , Q, T_H \rc$ characterize the black hole geometry and specify thermodynamic quantities of the dual QFT. The Bekenstein-Hawking entropy is determined by the area at the event horizon  
\be
S_{BH} = \fr{V_{d-1} }{4 G} \fr{R^{d-1}}{z_h^{d-1}} ,
\ee
where $G$ is a Newton constant of the bulk geometry and $V_{d-1}$ indicates a regularized volume at the boundary ($z=0$). In the holographic setup, the Bekenstein-Hawking entropy is identified with a thermal entropy of the dual QFT.

Now, let us take into account the entanglement entropy in a medium. For traditional QFTs, in general, calculating entanglement entropy in a medium is not easy. The holographic method in this situation can provide a new way to understand quantum entanglement. In the holographic setup, it was proposed that the entanglement entropy of QFT can be described by a minimal surface extending to the dual geometry. If we consider a medium consisting of massless radiation and massive objects, the entanglement entropy of this medium can be determined by a minimal surface extending to the previous $p$-brane gas geometry.  The entanglement entropy is given by a function of a subsystem size and implicitly depends on the densities of matters. After dividing the system into a strip-shaped subsystem and its complement, we parameterize a subsystem as $-\ell/2 \le x_1= x  \le \ell/2$ and $- L /2  \le x_i \le L/2$ with $ 0 \le \ell \le L$ for $i=2,\cdots, d-1$, where $L$ is a regularized length of the subsystem. Then, the entanglement entropy is governed by
\be
S_E = \fr{R^{d-1} L^{d-2}}{4 G} \int_{-\ell/2}^{\ell/2} dx \ \fr{ \sqrt{f + z'^2} }{z^{d-1} \sqrt{f}}  ,
\ee
where the prime means a derivative with respect to $x$. Using the conserved quantity, the subsystem size and entanglement entropy are determined as functions of a turning point $z_t$, at which $z'=0$, 
\be
\ell ( z_t ) &=& \int_\e^{z_t} dz \, \fr{2 z^{d-1}}{ \sqrt{f} \, \sqrt{z_t^{2 (d-1)} - z^{2 (d-1)} }} , \nn
S_E (z_t ) &=& \fr{R^{d-1} L^{d-2}}{2 G} \int_\e^{z_t}  dz \, \fr{z_t^{d-1}}{z^{d-1} \sqrt{f} \,  \sqrt{z_t^{2 (d-1)} - z^{2 (d-1)} } }  ,  \la{Relation:metee}
\ee
where $\e$ indicates a UV cutoff in the radial direction. After performing these integrals, we can determine the entanglement entropy as a function of the subsystem size, $S_E (\ell)$. In this case, since the metric factor $f$ relies on the density of matters, the resulting entanglement entropy must also depend on these matter's densities. If the metric factor $f(z)$ is known, one can holographically evaluate the entanglement entropy by applying the formulas in \eq{Relation:metee}. 

In the next section, we utilize \eq{Relation:metee} to make training sets for the reconstruction of the dual geometry. We first determine metric factors by randomly fixing the black hole parameters and then make training sets by evaluating their entanglement entropy. After training these data using the transformer method, we can inversely reconstruct the dual geometry of arbitrary entanglement entropy data, which we call a test set. It is worth noting that the holographic relation in \eq{Relation:metee}  is not crucial in this procedure. For example, when we have other training sets without applying the above holographic relation, training them allows us to find the dual geometry from a test set via the transformer model used in this work.

Before closing this section, there is an important remark. Using the Leibniz integral rule, we can find an important relation, the Hamilton-Jacobi equation, between entanglement entropy and the subsystem size \cite{Jokela:2021aa, Jokela:2023aa}
\begin{align}
	\frac{dS_E}{dl} = \frac{ L^{d-2} }{4G } \fr{R^{d-1}}{ z_t^{d-1}}  ,
	\label{dsdl}
\end{align}
where $L^{d-2}$ corresponds to the area of the strip's boundary. This implies that the known entanglement entropy $S(\ell)$ automatically determines the turning point as
\be
z_t =  \ls  \frac{ L^{d-2} }{4G } \fr{R^{d-1}}{d S_E/dl}  \rs^{1/(d-1)} .   \la{Relation:tpoint}
\ee
This relation enables us to reexpress the entanglement entropy as a function of the turning point, $S_E (z_t)$. In the holographic study, the position of the turning point can be reinterpreted as the inverse of the RG scale of the dual QFT \ct{Park:2024fik}. In the next section, we will reconstruct the dual geometry $f(z_t)$ from arbitrary entanglement entropy data by varying the subsystem size. In this case, the variation of the subsystem can be reinterpreted as the changes of the turning point due to the relation in \eq{Relation:tpoint}. As a result, this implies that one can reconstruct the dual geometry from the RG flow of entanglement entropy data. In addition, we get more information about other physical quantities from only the entanglement entropy data.

\subsection{Transformer AI}

\begin{figure}
\centering
\includegraphics[width=0.8\textwidth]{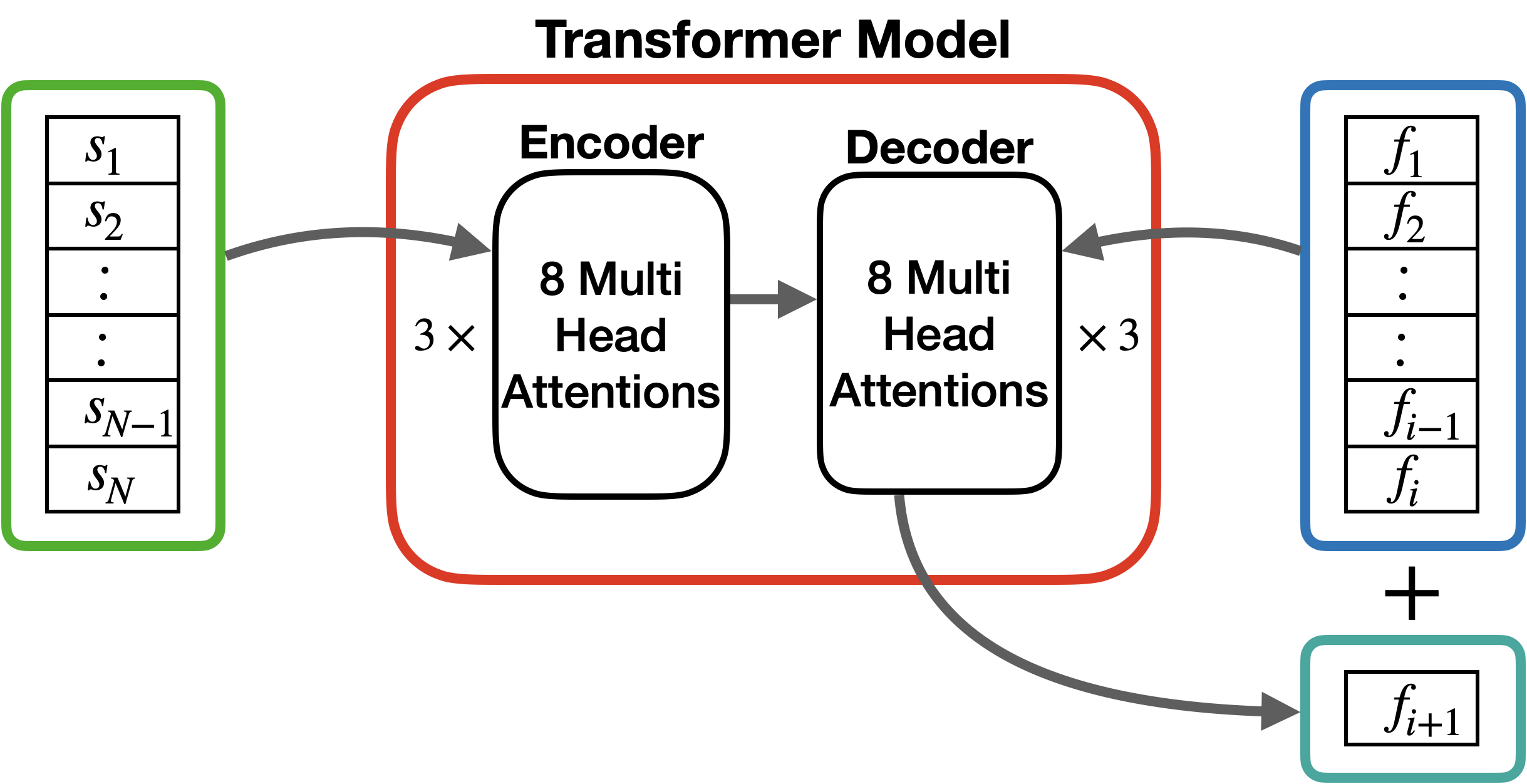}
\caption{This figure illustrates the structure of the transformer model used to extract information about holographic dual geometry from entanglement entropy data.}
\label{figure}
\end{figure}

The generative transformer model is a neural network architecture commonly employed in natural language processing tasks such as language translation and text summarization. It was first introduced in Ref.  \cite{Vaswani:2017aa} and is now widely utilized in various natural language applications \cite{Brown:2020aa, Devlin:2018aa, Touvron:2023aa, openai2023gpt4}. Due to the powerful capability of the transformer model, they are not limited to natural language processing alone but applied to computer vision and scientific research  \cite{Dosovitskiy:2020aa, Qin:2023aa, Bi:2023aa}.

The transformer model consists of two main components, the encoder and decoder blocks. The primary role of the encoder block is to comprehend information from the input sequence, while the decoder block generates the subsequent output sequence based on the encoded information. Both the encoder and decoder blocks include a multi-head self-attention layer. This layer takes three inputs, query, key, and value. It calculates the similarity between the query and key to determine their relationship, enabling the transformer to generate coherent output. In this paper, we construct a generative transformer model, as shown in Fig. \ref{figure}, to predict the dual geometry from holographic entanglement entropy data. Our model consists of three decoders and three encoders. Here we take 8 heads of attention layer for both the encoder and decoder. In Fig.  \ref{figure}, the input data in the green box means an input sequence of the entanglement entropy depending on the subsystem size $\ell$, while the blue box is an output sequence for the dual geometry related to the input data
\begin{align}
\lc z_i = \left. z_t \right|_{\ell=\ell_i} , \ s_i = S_E (z_i) \rc \quad \longrightarrow  \quad f_i = f(z_i) .
\label{data}
\end{align}
with
\be
\ell_i = \e + i \ \D \ell \quad {\rm with} \quad \D \ell = \fr{L-\e}{N} ,
\ee
where $N$ is a large integer number. The transformer model is a generative artificial intelligence model. It takes input and output sequence and generates the next output sequence, $f_{n+1}$ in Fig.  \ref{figure}, 
\begin{align}
{\rm \bf Transformer \ : }  \quad \sum_{i=0}^n s_i   \bigotimes  \sum_{i=0}^n f_i   \rightarrow  f_{n+1}.
\end{align}
To train the generative model, we used cross-entropy loss for the criterion and the Adam method as an optimizer. Once the transformer model is trained, we utilize it to predict the dual geometry of the given entanglement entropy. From this reconstructed dual geometry, we can also read the parameters, $\lc \r_p, Q, T_H \rc$, which give us information about the thermal properties of the considered system. As a result, reconstructing the dual geometry allows us to figure out the physical properties of the system, $\lc \r_p, Q, T_H \rc$, from only the entanglement entropy data, $S_E (\ell)$   

Before closing this section, we have some remarks on the transformer models for comparison with deep learning models:

\begin{itemize}

   \item In Ref. \cite{Park:2022fqy}, deep learning models have to cooperate with the holographic formula \eq{Relation:metee} to reconstruct the dual geometry. For transformer models, however, we do not need to priorly know how to obtain training big data. If we have a sufficient number of training big data independent of the holographic principle, the transformer model extract the hidden underlying structure between data.
   
  \item Deep learning methods require successive optimization to obtain the dual geometry satisfying the holographic principle, while transformer models directly give rise to the dual geometry determined by the hidden relation derived from training sets. 
  
  \item Even after removing the UV divergence, the entanglement entropy of thermal systems still suffers from IR divergence appearing at the event horizon, which in the numerical work causes a large numerical error. Due to this problem, it is not easy to determine the event horizon exactly in deep learning models. In contrast, transformer models can avoid this large numerical error and allow us to determine the event horizon more accurately.
  
  \item In \cite{Hashimoto:2018wm, Park:2022fqy}, deep learning models consist of specific neural layers related to the holographic principle. Therefore, one can understand the mechanism working in deep learning modes. However, the transformer model is an artificial neural network model, whose inner structure is obscure. In this work, we follow the guideline in Ref. \cite{Popel_2018} to train the transformer model.

\end{itemize}


\section{Reconstructing dual geometries of entanglement entropy}

In this section, we discuss how to reconstruct a three-dimensional black hole geometry explicitly from arbitrary two-dimensional entanglement entropy data. To do so, let us consider a two-dimensional thermal system composed of massless radiation and two massive particles. The dual of this thermal system can be described by the following gravity action
\be
S = \fr{1}{16 \pi G} \int d^3 x \sqrt{-g} \ls {\cal R} - 2  \L - F_{\m\n} F^{\m\n} \rs +  T  N \int d^{2} \xi \sqrt{- h} h^{\a \b} \pa_\a x^\m \pa_\b x^\n g_{\m\n} ,
\ee 
where $T$ and $N$ indicate the tension and number of $1$-branes and $h_{\a\b}$ is an induced metric on $1$-branes. If $D1$-branes extend to the radial direction, they can be reinterpreted as heavily massive particles like solitons on the dual QFT side. On the other hand, the bulk gauge field $F_{\m\n}$ is mapped to lightly massive particles like hadrons in the holographic QCD \ct{Lee:2009bya,Park:2009nb}. This feature becomes manifest in a geometric solution \ct{Park:2022abi,Park:2021wep}. Solving the Euler equations, we find a black hole-type geometric solution
\be
ds^2 = \fr{R^2}{z^2} \ls - f (z) dt^2 + \fr{1}{f(z)} dz^2 + dx^2   \rs ,
\ee
with the following blackening factor
\be
f (z) = 1 -  \r \, z - M z^2 + \fr{Q^2}{2} z^2 \log \fr{z}{R} ,
\ee
where $\r$, $M$, and $Q$ are related to the density of heavily massive particles, massless radiations, and lightly massive particles, respectively. Above, the logarithmic term appears due to the degeneracy of the black hole mass and gauge field in a three-dimensional AdS space. Recalling that the black hole mass can be rewritten as a function of the other parameters
 \be
 M = \fr{1}{z_h^2} - \fr{\r}{z_h} + \fr{1}{2} Q^2 \log \fr{z_h}{R}  ,
 \ee
 where $z_h$ is an event horizon, the black hole solution is specified by three parameters or hairs, $\lc \r , Q , T_H\rc$. In the holographic setup, these three parameters characterize the dual thermal system. For example, the temperature of a considered thermal system is given by the Hawking temperature 
\be
T_H = - \lp \fr{f'(z)}{4 \pi } \right|_{z=z_h}  = \fr{1}{2 \pi z_h} - \fr{\r}{4 \pi} - \fr{Q^2 z_h}{8 \pi} .
\ee
In addition, the massive matters of this thermal system are specified by $\r$ and $Q$. 

When the entanglement entropy of this thermal system is known as a function of the subsystem size, can we determine the parameters, $\lc \r , Q , T_H\rc$, characterizing the thermal system? If the underlying theory is not known, there is no way to find the connection between physical quantities and entanglement entropy from the theoretical viewpoint. According to the AdS/CFT correspondence, however, it is possible to find the relation between physical parameters and entanglement entropy by reconstructing the dual geometry. This indicates that the reconstructed dual geometry can play the role of an effective theory governing physical quantities. In the present work, we discuss how to determine the physical parameters, $\lc \r , Q , T_H\rc$, from entanglement entropy, $S(\ell)$, by reconstructing the dual geometry via the transformer model. After training the relation between the metric factor and entanglement entropy in the transformer model, we will derive the physical quantities characterizing a thermal system from arbitrary entanglement entropy data.

\subsection{How to make training sets}

Now, we consider the entanglement entropy of a medium consisting of massless radiations and massive particles. On the QFT side, it is very difficult to evaluate the entanglement entropy due to the nontrivial interaction between matters. Therefore, we calculate it on the dual gravity side by applying the holographic proposal where the quantum entanglement entropy is described by the area of a minimal surface extending to the dual geometry. To define the entanglement entropy in the dual QFT, we divide the boundary space into two parts, a subsystem with a size $\ell$ and its complement. On the gravity side, then, a geodesic curve anchored at the entangling surface describes the entanglement entropy, which on the dual gravity side is governed by
\be
S_E (\ell) = \fr{R}{4 G}   \int_{-\ell/2}^{\ell/2} dx  \, \fr{ \sqrt{f + z'^2 } }{z  \, \sqrt{ f}}  ,
\ee 
where the prime means a derivative with respect to $x$ and the blackening factor $f$ becomes in terms of $\lc \r , Q, z_h \rc$
\be
f = 1 -  \r \, z - \ls \fr{1}{z_h^2} - \fr{\r}{z_h} + \fr{1}{2} Q^2 \log \fr{z_h}{R} \rs z^2 + \fr{Q^2}{2} z^2 \log \fr{z}{R} ,
\ee 
Introducing a turning point $z_t$ where $z'=0$, the subsystem size and entanglement entropy are represented as functions of the turning point
\be
\ell (z_t) &=& \int_0^{z_t} dz \fr{2 z }{\sqrt{f}  \, \sqrt{z_t^2 -z^2}} , \nn
S_E (z_t) &=&\fr{R  }{2 G}  \int_\e^{z_t} dz \fr{z_t }{ z  \sqrt{f}  \, \sqrt{z_t^2 -z^2}} .  \la{Relation:lSE}
\ee
When $\e \to 0$, the entanglement entropy has a logarithmic divergence. To avoid this UV divergence issue, we introduce a physical UV cutoff $\e$ like a lattice spacing, which is small but finite. Performing these integrals numerically, the subsystem size and entanglement entropy are determined as functions of the turning point. Removing the turning point by combining $\ell (z_t)$ and $S_E (z_t)$, moreover, we can also represent the entanglement entropy as a function of the subsystem size, $S_E (\ell)$. In this calculation, we can take into account either $z_t$ or $\ell$ as a free parameter due to \eq{dsdl}. 

Since the three hairs in the present model specify the black hole geometry, we can make many training sets by considering various sets of hairs. We make $100,000$ training sets from $100,000$ combinations of three hairs, $\lc \r , M , Q \rc$. Although we consider only three hairs for convenience, the present work can be further generalized into general cases with more hairs in an arbitrary dimension. When training sets without knowing how to make them are given, the relation between input and output data is not manifest. However, the transformer method tries to find a multiple matrix transformation between inputs to outputs from the given training data. This transformation determines the relation between input and output data implicitly. When an arbitrary test set, $\lc \ell, S_E (\ell) \rc$ or $\lc z_t , S_E (z_t) \rc$, is given, one can find a dual geometry $\lc z_t, f (z_t) \rc$, of the given test set with satisfying the matrix relation derived from the training set.

\subsection{Dual geometry of test sets}

To find physical quantities specifying the dual geometry from given entanglement entropy data, we first assume that the entanglement entropy of a thermal system is known as a function of the subsystem size, $\lc \ell, S_E (\ell)\rc$.  In this case, the subsystem size $\ell$ becomes a free parameter and the turning point $z_t$ can be determined from the given entanglement entropy data
\be
z_t = \fr{R}{4 G } \fr{1}{d S_E / d \ell} .
\ee
Using this relation, we can exploit $z_t$, instead of $\ell$, as a free parameter and represent the entanglement entropy as a function of $z_t$ instead of $\ell$. Then, we try to the dual geometry $\lc z_t, f(z_t) \rc$ from the entanglement entropy data, $\lc z_t , S_E (z_t) \rc$, by varying $z_t$.

We first consider a BTZ black hole with $\r=Q=0$ and make entanglement entropy data as test sets for $z_h=0.5$, $1.0$, and $1.5$. In Fig. 2(a), we plot the resulting entanglement entropy relying on the subsystem size. Assuming we do not know how to obtain these entanglement entropy data, we try reconstructing the dual geometry only from the given entanglement entropy data. Using the transformer model with the previous  $100,000$ training sets, we find the dual geometries of the given entanglement entropy data, which are depicted in Fig. 2(b). From the reconstructed dual geometries, we extract physical quantities in Table 1, which characterize the thermodynamic properties hidden in the entanglement entropy data in Fig. 2(a). These physical quantities, moreover, determine the temperature of the thermal system, which gives rise to the entanglement entropy data in Fig. 2(a). As a consequence, the obtained results show that the transformer model allows us to reconstruct the dual geometry from the entanglement entropy data and gives us information about physical quantities hidden in the entanglement entropy data with small numerical errors.

\begin{figure}
	\centering
	\begin{subfigure}[b]{0.45\textwidth}
		\centering
		\includegraphics[width=\textwidth]{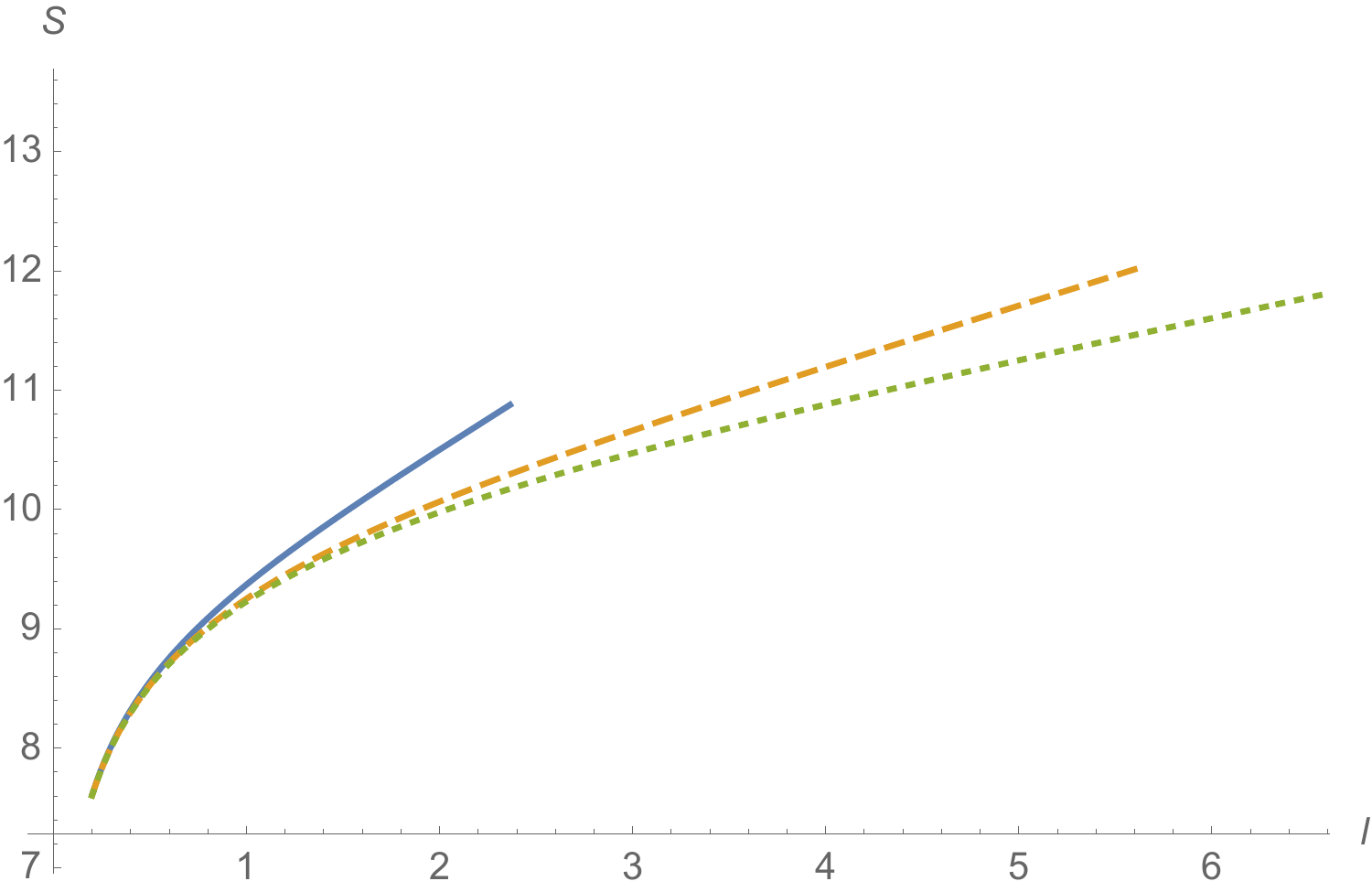}
		\caption{Entanglement entropy data for $\r=Q=0$}
	\end{subfigure}
	\hfill
	\begin{subfigure}[b]{0.45\textwidth}
		\centering
		\includegraphics[width=\textwidth]{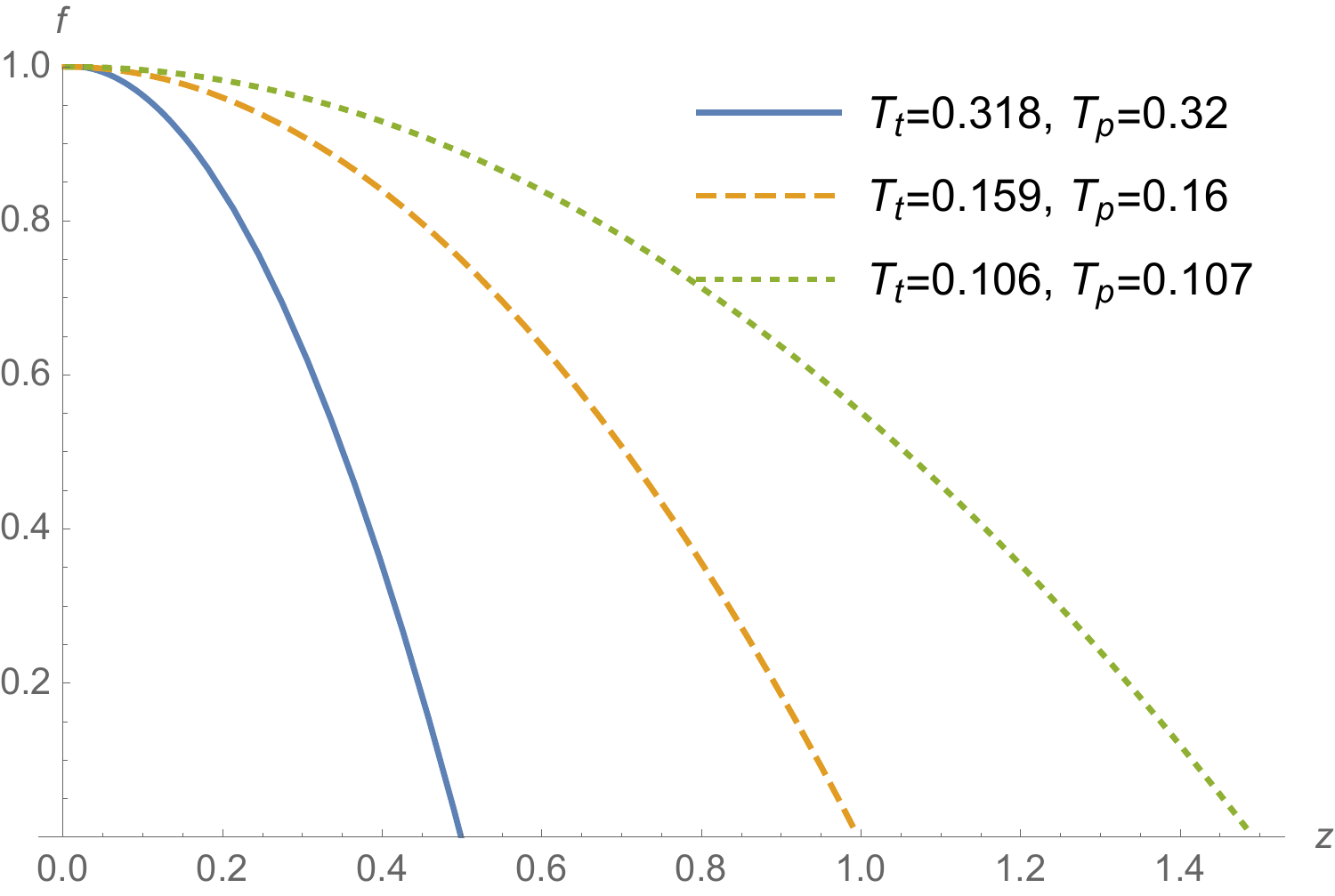}
		\caption{Prediction of the dual geometries}
	\end{subfigure}
	\caption{(a) For the BTZ black hole with $\r=Q=0$, we plot the entanglement entropy data with $z_h = 0.5$ (blue-solid curve), $1.0$ (orange-dashed curve), and $1.5$ (green-dotted curve). (b) From the entanglement entropy data given in (a), we reconstruct the dual geometries. The dashed lines present the prediction by the transformer model, whereas the solid lines indicate the true blackening factor we used. $T_t$ is the Hawking temperature derived from the true values, while $T_p$ is the predicted temperature by the transformer model. \\}
\end{figure} 

\begin{table}[h!]
\centering
\begin{tabular}{|c|c|c|c|c|c|c|}
\hline
 & \multicolumn{2}{c|}{\textbf{Blue-solid}} & \multicolumn{2}{c|}{\textbf{Orange-dashed}} & \multicolumn{2}{c|}{\textbf{Green-dotted}} \\ \hline
\multicolumn{1}{|c|}{\textbf{Parameter}}                                          & \textbf{Pred}            & \textbf{True}            & \textbf{Pred }            & \textbf{True }            & \textbf{Pred }            & \textbf{True }            \\ \hline
$z_h$                             & 0.4976                      & 0.5                     & 0.9968                      & 1.0                     & 1.4916                      & 1.5                     \\ \hline
$\rho$                                & 0.0012                      & 0.0                     & -0.0022                      & 0.0                     & -0.0027                      & 0.0                     \\ \hline
$Q$                              & 0.000                      & 0.0                     & 0.000                      & 0.0                     & 0.0000                      & 0.0                     \\ \hline
$T$                              & 0.32                      & 0.318                    & 0.16                      & 0.159                     & 0.107                      & 0.106                     \\ \hline
\end{tabular}
\caption{The physical quantities derived from the reconstructed dual geometries in Fig. 2(b). }
\label{tab:comparison}
\end{table}

\begin{figure}
  \centering
	\begin{subfigure}[b]{0.45\textwidth}
	\centering
	\includegraphics[width=\textwidth]{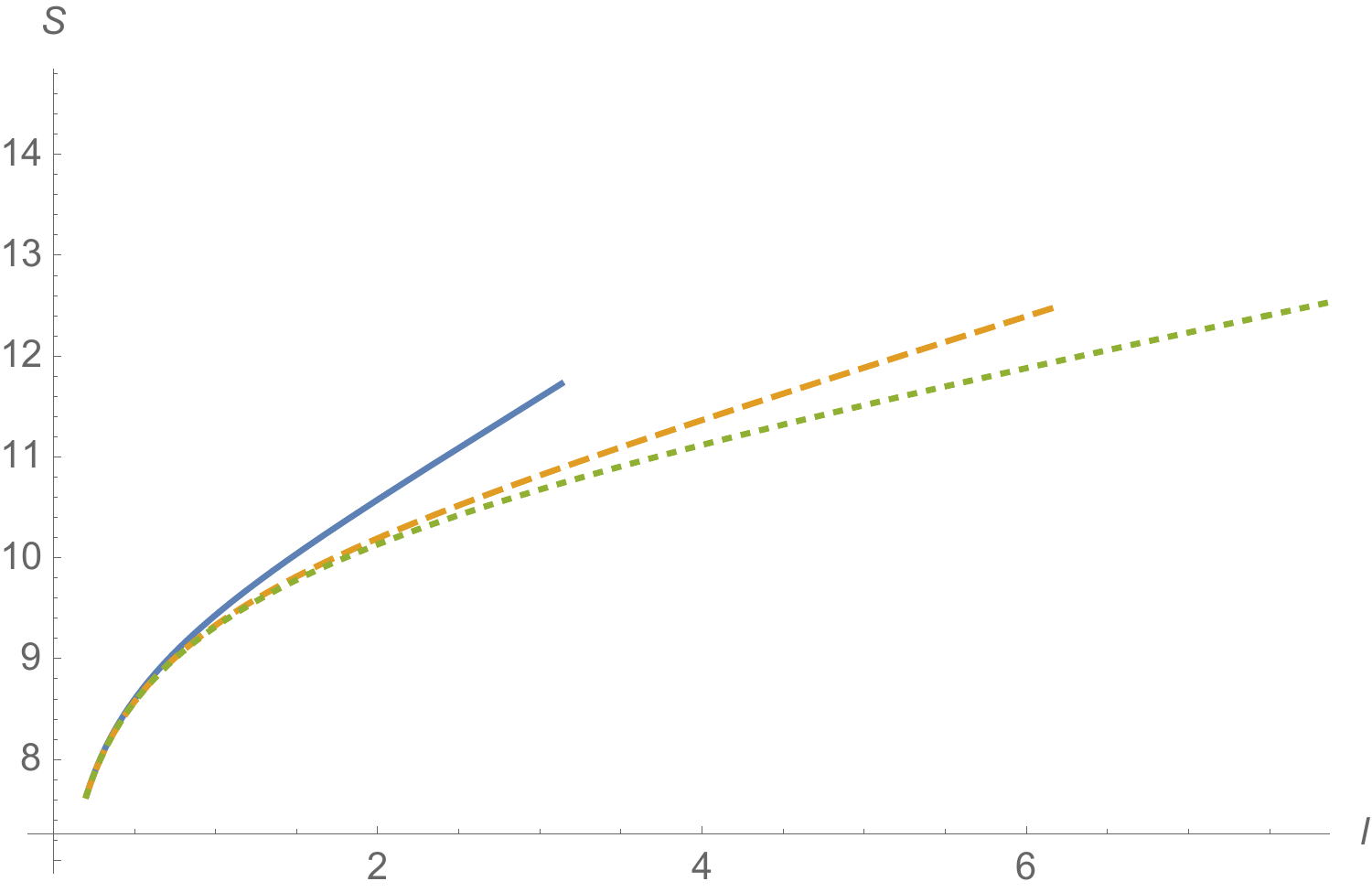}
	\caption{Entanglement entropy data for  $\rho=0.5$ and $Q=0$}
	\end{subfigure}
	\hfill
	\begin{subfigure}[b]{0.45\textwidth}
	\centering
	\includegraphics[width=\textwidth]{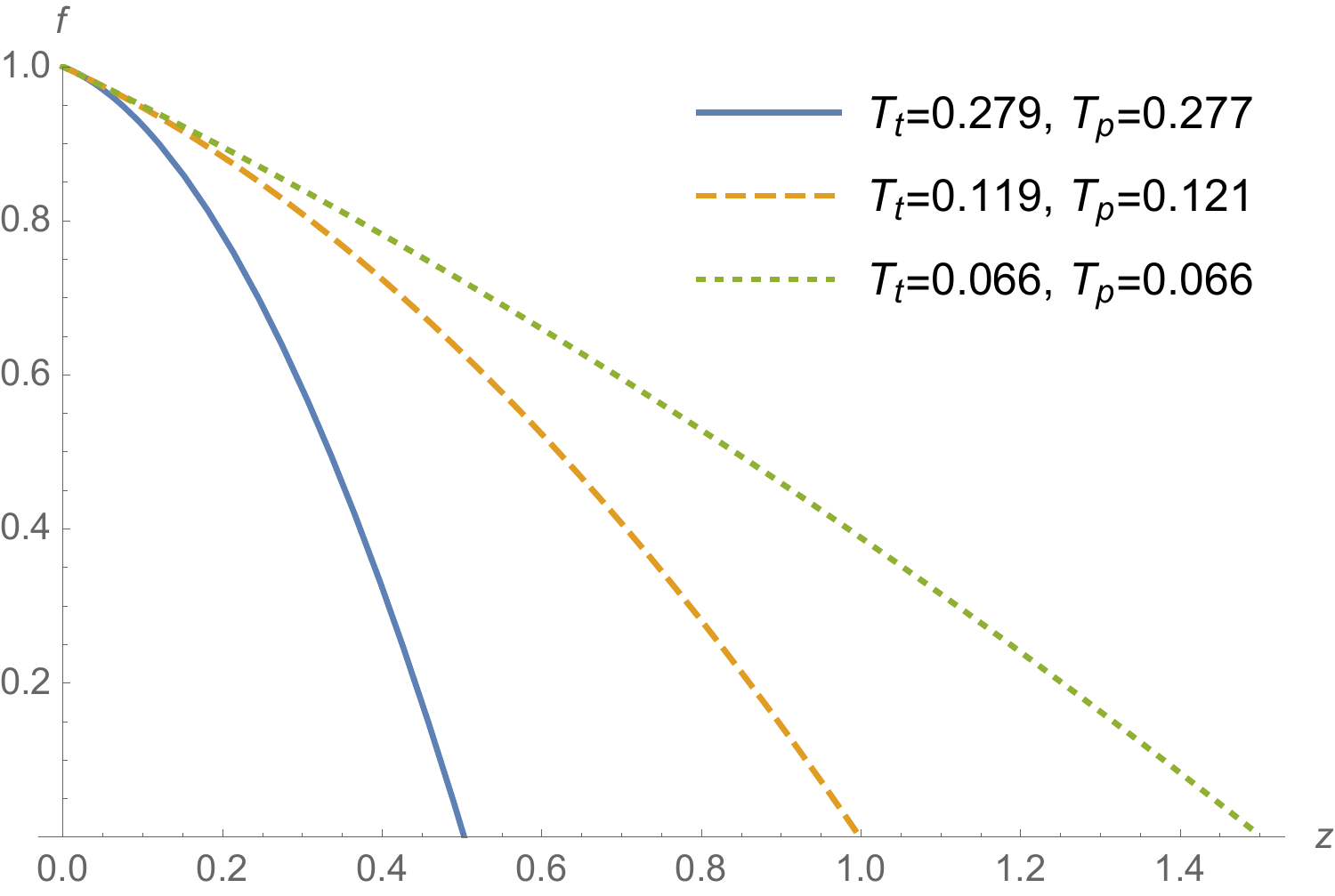}
	\caption{Predicted dual geometries for $\rho=0.5$ and $Q=0$ }
	\end{subfigure}
\centering
	\begin{subfigure}[b]{0.45\textwidth}
	\centering
	\includegraphics[width=\textwidth]{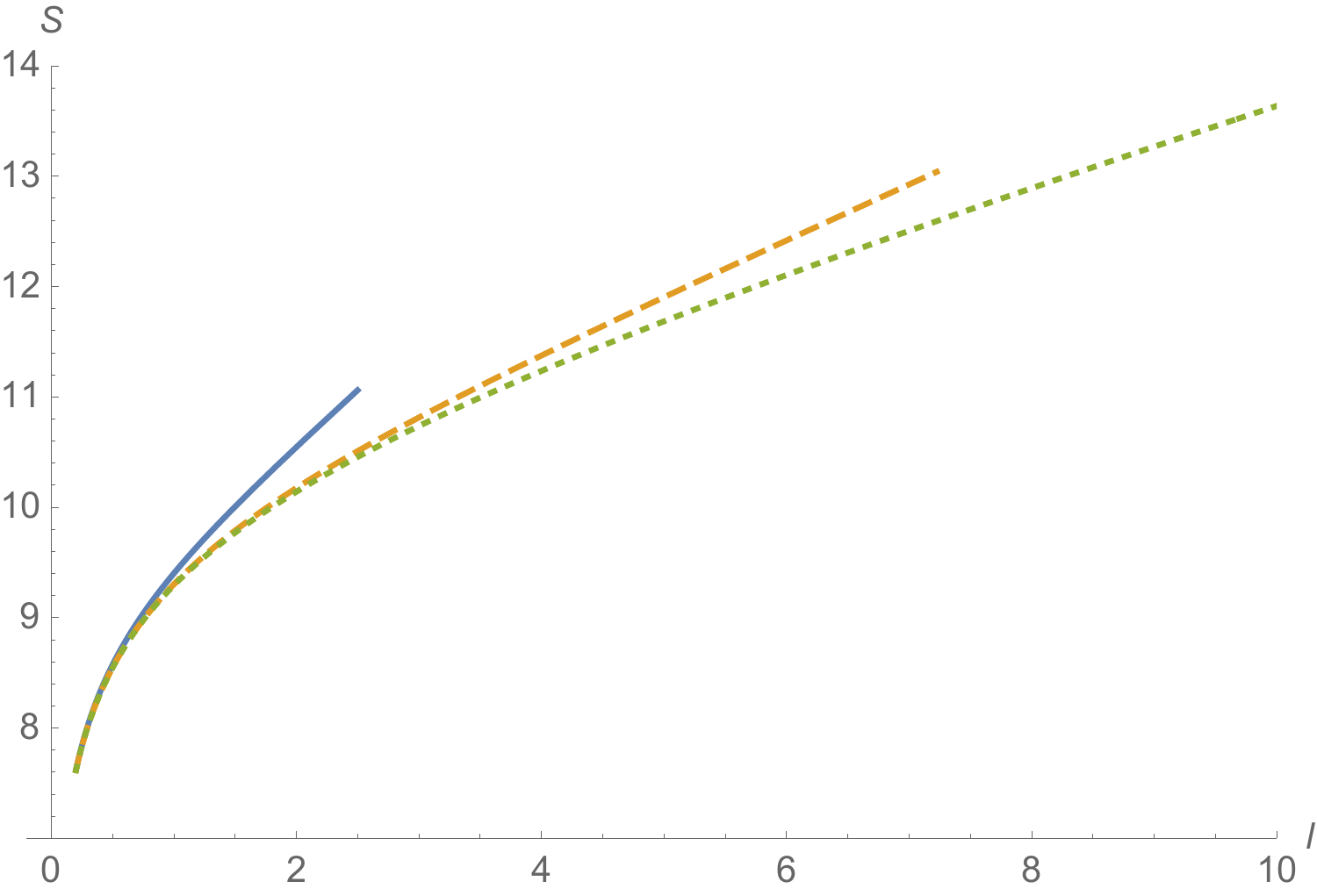}
	\caption{Entanglement entropy data for $\rho=0$ and $Q=0.65$}
	\end{subfigure}
	\hfill
	\begin{subfigure}[b]{0.45\textwidth}
	\centering
	\includegraphics[width=\textwidth]{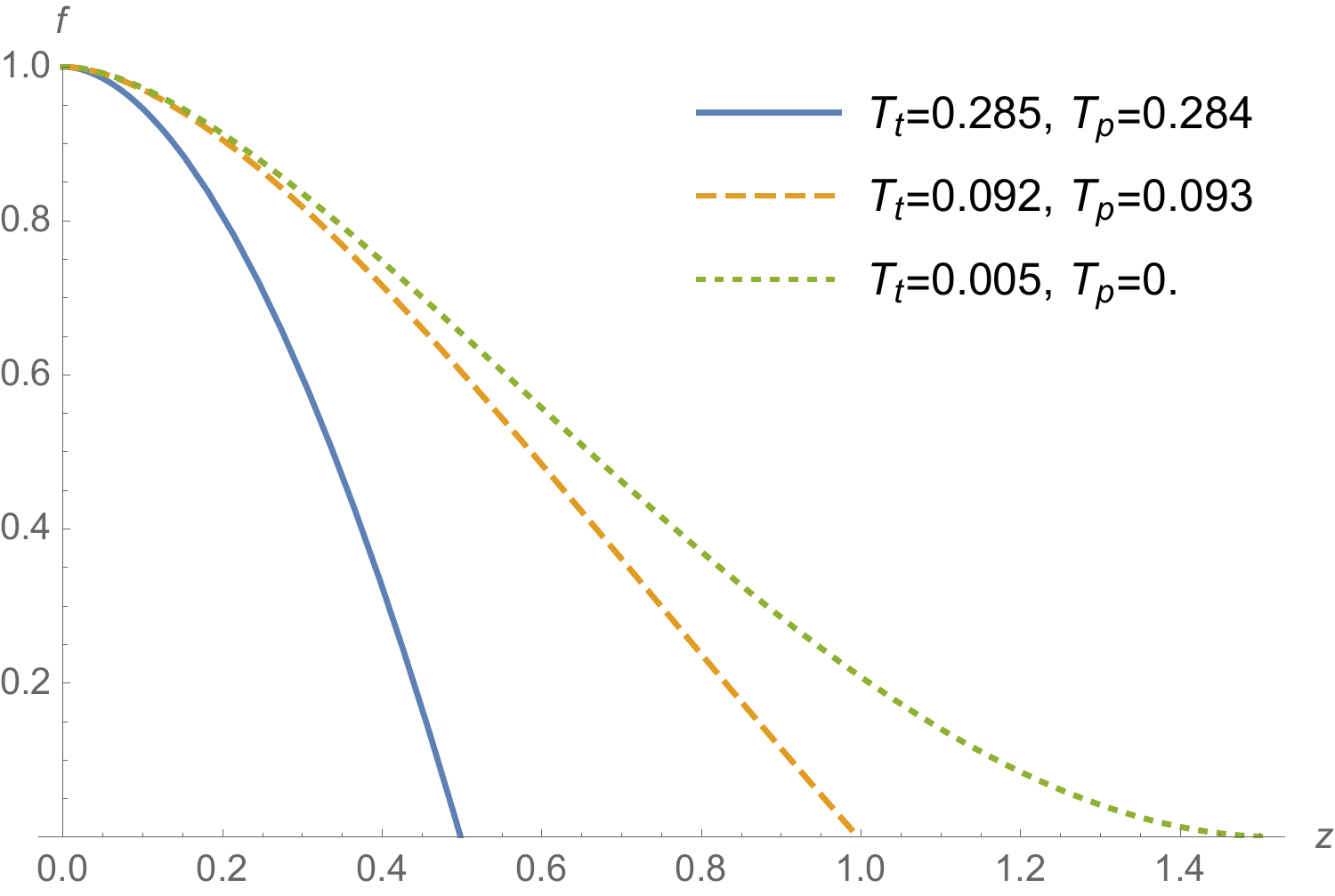}
	\caption{Predicted dual geometries for $\rho=0$ and $Q=0.65$ }
	\end{subfigure}
\centering
	\begin{subfigure}[b]{0.45\textwidth}
	\centering
	\includegraphics[width=\textwidth]{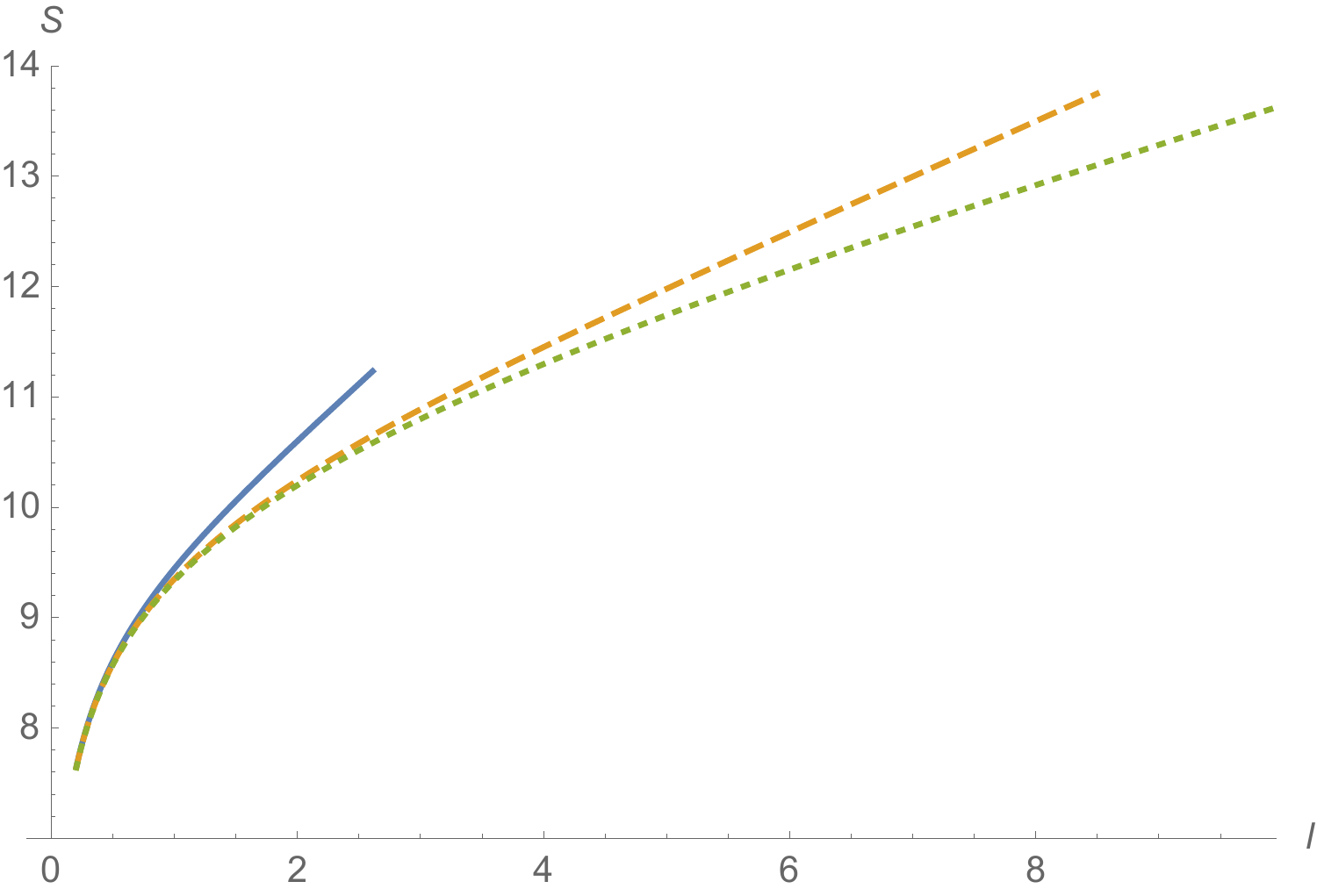}
	\caption{Entanglement entropy data for $\rho=0.4$ and $Q=0.5$  }
	\end{subfigure}
	\hfill
	\begin{subfigure}[b]{0.45\textwidth}
	\centering
	\includegraphics[width=\textwidth]{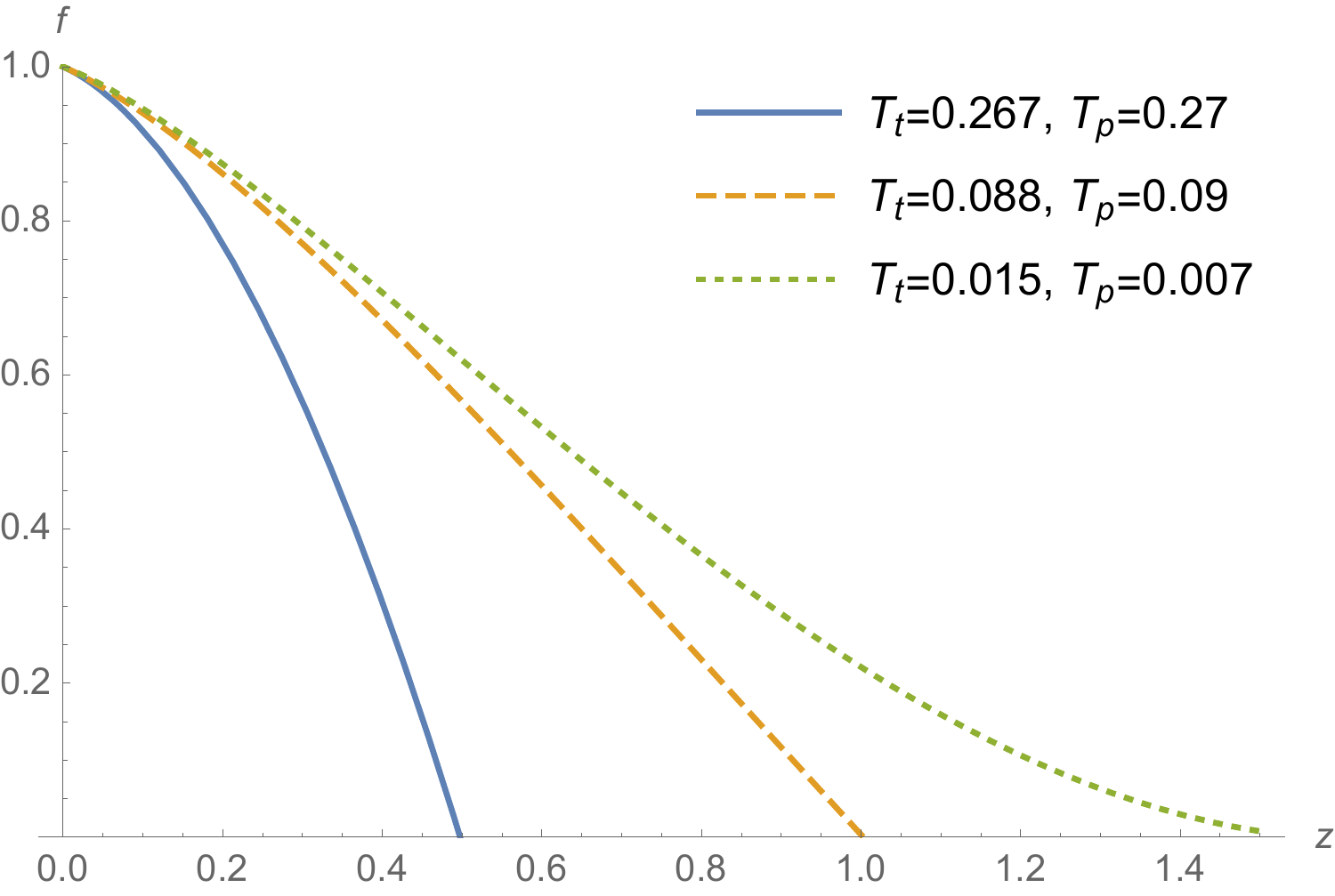}
	\caption{Predicted dual geometries for $\rho=0.4$ and $Q=0.5$ }
	\end{subfigure}
\caption{The entanglement entropy data (left figures) and their dual geometries (right figures) reconstructed by the transformer model, where we use $z_h = 0.5$ (blue-solid curve), $1.0$ (orange-dashed curve), and $1.5$ (green-dotted curve).}
\end{figure}

Now, let us take into account more general cases. We first make the entanglement entropy data, Fig. 3(a) for $\r=0.5$ and $Q=0$, Fig. 3(c) for $\r=0$ and $Q=0.65$, and Fig. 3(e) for $\r=0.4$ and $Q=0.5$. Even when we do not know how to make these entanglement entropy data, we can reconstruct their dual geometries by the transformer model. The resulting blackening factors are represented in Fig. 3(b) for $\r=0.5$ and $Q=0$, Fig. 3(d) for $\r=0$ and $Q=0.65$, and Fig. 3(f) for $\r=0.4$ and $Q=0.5$. The obtained results show that the transformer method can reconstruct the dual geometries consistent with the true ones up to small numerical errors. Moreover, the reconstructed dual geometries allow us to find the physical quantities characterizing the thermal system. In Table 2 for  $\r=0.5$ and $Q=0$, Table 3 for $\r=0$ and $Q=0.65$, and Table 4 for $\r=0.4$ and $Q=0.5$, we represent the physical quantities expected from the reconstructed dual geometries, which are consistent with the true values up to small numerical errors. These physical quantities also determine the system's temperature consistently, as shwon in Fig. 3, with small numerical errors.

\begin{table}[h!]
\centering
\begin{tabular}{|c|c|c|c|c|c|c|}
\hline
 & \multicolumn{2}{c|}{\textbf{Blue-solid}} & \multicolumn{2}{c|}{\textbf{Orange-dashed}} & \multicolumn{2}{c|}{\textbf{Green-dotted}} \\ \hline
\multicolumn{1}{|c|}{\textbf{Parameter}}                                          & \textbf{Pred}            & \textbf{True}            & \textbf{Pred }            & \textbf{True }            & \textbf{Pred }            & \textbf{True }            \\ \hline
$z_h$                             & 0.5019                      & 0.5                     & 0.9987                      & 1.0                     & 1.5022                      & 1.5                     \\ \hline
$\rho$                                & 0.4790                      & 0.5                     & 0.4744                      & 0.5                     & 0.4963                      & 0.5                     \\ \hline
$Q$                              & 0.1391                      & 0.0                     & 0.0774                      & 0.0                     & 0.0538                      & 0.0                     \\ \hline
$T$                              & 0.277                      & 0.279                    & 0.121                      & 0.119                     & 0.066                      & 0.066                     \\ \hline
\end{tabular}
\caption{The physical quantities derived from the reconstructed dual geometries in Fig. 3(b)  }
\label{tab:comparison}
\end{table}

\begin{table}[h!]
\centering
\begin{tabular}{|c|c|c|c|c|c|c|}
\hline
 & \multicolumn{2}{c|}{\textbf{Blue-solid}} & \multicolumn{2}{c|}{\textbf{Orange-dashed}} & \multicolumn{2}{c|}{\textbf{Green-dotted}} \\ \hline
\multicolumn{1}{|c|}{\textbf{Parameter}}                                          & \textbf{Pred}            & \textbf{True}            & \textbf{Pred }            & \textbf{True }            & \textbf{Pred }            & \textbf{True }            \\ \hline
$z_h$                             & 0.4984                      & 0.5                     & 0.9964                      & 1.0                     & 1.5264                      & 1.5                     \\ \hline
$\rho$                                & 0.0009                      & 0.0                     & 0.0052                      & 0.0                     & -0.0004                      & 0.0                     \\ \hline
$Q$                              & 0.6634                      & 0.65                     & 0.6460                      & 0.65                     & 0.6552                      & 0.65                     \\ \hline
$T$                              & 0.284                      & 0.285                    & 0.093                      & 0.092                     & 0.0                      & 0.005                     \\ \hline
\end{tabular}
\caption{The physical quantities derived from the reconstructed dual geometries in Fig. 3(d)   }
\label{tab:comparison}
\end{table}

\begin{table}[h!]
\centering
\begin{tabular}{|c|c|c|c|c|c|c|}
\hline
& \multicolumn{2}{c|}{\textbf{Blue-solid}} & \multicolumn{2}{c|}{\textbf{Orange-dashed}} & \multicolumn{2}{c|}{\textbf{Green-dotted}} \\ \hline
\multicolumn{1}{|c|}{\textbf{Parameter}}       & \textbf{Pred}            & \textbf{True}            & \textbf{Pred }            & \textbf{True }            & \textbf{Pred }            & \textbf{True }            \\ \hline
$z_h$                             & 0.4974                      & 0.5                     & 1.0029                      & 1.0                     & 1.5607                      & 1.5                     \\ \hline
$\rho$                                & 0.4757                      & 0.4                     & 0.4504                      & 0.4                     & 0.3857                      & 0.4                     \\ \hline
$Q$                              & 0.3926                      & 0.5                     & 0.4567                      & 0.5                     & 0.5084                      & 0.5                     \\ \hline
$T$                              & 0.27                      & 0.267                    & 0.09                      & 0.088                     & 0.007                      & 0.0015                     \\ \hline
\end{tabular}
\caption{The physical quantities derived from the reconstructed dual geometries in Fig. 3(f)  }
\label{tab:comparison}
\end{table}


\section{Discussion}

We have studied the transformative application of a transformer model to the holographic reconstruction of dual geometries from entanglement entropy data. For convenience, we focused only on the entanglement entropy of two-dimensional thermal systems characterized by three thermodynamic quantities, but the method can easily be applied to higher-dimensional cases. 

After training many sets of entanglement entropy and the corresponding dual geometry, the transformer model can learn the implicit relation working in the training sets. Using this relation, we reconstructed the dual geometry of arbitrary entanglement entropy as a test set. In this procedure, a test set contains information about entanglement entropy relying only on the subsystem size. Therefore, we cannot directly read thermodynamic quantities hidden in entanglement entropy data if we do not know the underlying theory governing entanglement entropy and thermodynamic quantities. However, if we can reconstruct the dual gravity theory of entanglement entropy data, we can get information about thermodynamic quantities from the entanglement entropy. This is because the dual gravity theory can play the role of an effective theory governing the relation between entanglement entropy and thermodynamic quantities. In the present work, we derived the thermodynamic quantities from the entanglement entropy data depending only on the subsystem size with small numerical errors.

In this work, we employed only the entanglement entropy to test the transformative model. However, quantum information theory has a variety of informational quantities measuring many different quantum features. For example, mutual information measures the correlation between two subregions and is defined as the combination of entanglement entropies
\be
\text{MI}(A;B)=S(A)+S(B)-S(A\cup B) .
\ee 
This mutual information captures the phase change between connected and disconnected minimal surfaces. Our model can be utilized to study this kind of first-order phase transition and extend to predict specific geometric solutions related to the second-order phase transition. We leave these works as future direction.

\vspace{1cm}


{\bf Acknowledgement}

C. Park was supported by the National Research Foundation of Korea (NRF) grant funded by the Korean government (No. NRF-2019R1A2C1006639). J. H. Lee was supported by the National Research Foundation of Korea (NRF) grant funded by the Korean government (No. NRF-2021R1C1C2008737). S. Kim was supported by the Basic Science Research Program through NRF grant (No. NRF-2016R1D1A1B03931443).


\bibliographystyle{apsrev4-1.bst}
\bibliography{ref}

\begin{thebibliography}{48}%
\makeatletter
\providecommand \@ifxundefined [1]{%
 \@ifx{#1\undefined}
}%
\providecommand \@ifnum [1]{%
 \ifnum #1\expandafter \@firstoftwo
 \else \expandafter \@secondoftwo
 \fi
}%
\providecommand \@ifx [1]{%
 \ifx #1\expandafter \@firstoftwo
 \else \expandafter \@secondoftwo
 \fi
}%
\providecommand \natexlab [1]{#1}%
\providecommand \enquote  [1]{``#1''}%
\providecommand \bibnamefont  [1]{#1}%
\providecommand \bibfnamefont [1]{#1}%
\providecommand \citenamefont [1]{#1}%
\providecommand \href@noop [0]{\@secondoftwo}%
\providecommand \href [0]{\begingroup \@sanitize@url \@href}%
\providecommand \@href[1]{\@@startlink{#1}\@@href}%
\providecommand \@@href[1]{\endgroup#1\@@endlink}%
\providecommand \@sanitize@url [0]{\catcode `\\12\catcode `\$12\catcode
  `\&12\catcode `\#12\catcode `\^12\catcode `\_12\catcode `\%12\relax}%
\providecommand \@@startlink[1]{}%
\providecommand \@@endlink[0]{}%
\providecommand \url  [0]{\begingroup\@sanitize@url \@url }%
\providecommand \@url [1]{\endgroup\@href {#1}{\urlprefix }}%
\providecommand \urlprefix  [0]{URL }%
\providecommand \Eprint [0]{\href }%
\providecommand \doibase [0]{http://dx.doi.org/}%
\providecommand \selectlanguage [0]{\@gobble}%
\providecommand \bibinfo  [0]{\@secondoftwo}%
\providecommand \bibfield  [0]{\@secondoftwo}%
\providecommand \translation [1]{[#1]}%
\providecommand \BibitemOpen [0]{}%
\providecommand \bibitemStop [0]{}%
\providecommand \bibitemNoStop [0]{.\EOS\space}%
\providecommand \EOS [0]{\spacefactor3000\relax}%
\providecommand \BibitemShut  [1]{\csname bibitem#1\endcsname}%
\let\auto@bib@innerbib\@empty
\bibitem [{\citenamefont {Maldacena}(1999)}]{Maldacena:1997re}%
  \BibitemOpen
  \bibfield  {author} {\bibinfo {author} {\bibfnamefont {J.~M.}\ \bibnamefont
  {Maldacena}},\ }\href {\doibase 10.1023/A:1026654312961,
  10.4310/ATMP.1998.v2.n2.a1} {\bibfield  {journal} {\bibinfo  {journal} {Int.
  J. Theor. Phys.}\ }\textbf {\bibinfo {volume} {38}},\ \bibinfo {pages} {1113}
  (\bibinfo {year} {1999})},\ \bibinfo {note} {[Adv. Theor. Math.
  Phys.2,231(1998)]},\ \Eprint {http://arxiv.org/abs/hep-th/9711200}
  {arXiv:hep-th/9711200 [hep-th]} \BibitemShut {NoStop}%
\bibitem [{\citenamefont {Gubser}\ \emph {et~al.}(1998)\citenamefont {Gubser},
  \citenamefont {Klebanov},\ and\ \citenamefont {Polyakov}}]{Gubser:1998bc}%
  \BibitemOpen
  \bibfield  {author} {\bibinfo {author} {\bibfnamefont {S.~S.}\ \bibnamefont
  {Gubser}}, \bibinfo {author} {\bibfnamefont {I.~R.}\ \bibnamefont
  {Klebanov}}, \ and\ \bibinfo {author} {\bibfnamefont {A.~M.}\ \bibnamefont
  {Polyakov}},\ }\href {\doibase 10.1016/S0370-2693(98)00377-3} {\bibfield
  {journal} {\bibinfo  {journal} {Phys. Lett. B}\ }\textbf {\bibinfo {volume}
  {428}},\ \bibinfo {pages} {105} (\bibinfo {year} {1998})},\ \Eprint
  {http://arxiv.org/abs/hep-th/9802109} {arXiv:hep-th/9802109} \BibitemShut
  {NoStop}%
\bibitem [{\citenamefont {Witten}(1998{\natexlab{a}})}]{Witten:1998aa}%
  \BibitemOpen
  \bibfield  {author} {\bibinfo {author} {\bibfnamefont {E.}~\bibnamefont
  {Witten}},\ }\href {https://arxiv.org/pdf/hep-th/9802150.pdf} {\bibfield
  {journal} {\bibinfo  {journal} {Adv.Theor.Math.Phys.}\ }\textbf {\bibinfo
  {volume} {2}},\ \bibinfo {pages} {253} (\bibinfo {year}
  {1998}{\natexlab{a}})},\ \Eprint {http://arxiv.org/abs/hep-th/9802150}
  {hep-th/9802150} \BibitemShut {NoStop}%
\bibitem [{\citenamefont {Witten}(1998{\natexlab{b}})}]{Witten:1998ab}%
  \BibitemOpen
  \bibfield  {author} {\bibinfo {author} {\bibfnamefont {E.}~\bibnamefont
  {Witten}},\ }\href {https://arxiv.org/pdf/hep-th/9803131.pdf} {\bibfield
  {journal} {\bibinfo  {journal} {Adv.Theor.Math.Phys.}\ }\textbf {\bibinfo
  {volume} {2}},\ \bibinfo {pages} {505} (\bibinfo {year}
  {1998}{\natexlab{b}})},\ \Eprint {http://arxiv.org/abs/hep-th/9803131}
  {hep-th/9803131} \BibitemShut {NoStop}%
\bibitem [{\citenamefont {Aharony}\ \emph {et~al.}(2000)\citenamefont
  {Aharony}, \citenamefont {Gubser}, \citenamefont {Maldacena}, \citenamefont
  {Ooguri},\ and\ \citenamefont {Oz}}]{Aharony:1999ti}%
  \BibitemOpen
  \bibfield  {author} {\bibinfo {author} {\bibfnamefont {O.}~\bibnamefont
  {Aharony}}, \bibinfo {author} {\bibfnamefont {S.~S.}\ \bibnamefont {Gubser}},
  \bibinfo {author} {\bibfnamefont {J.~M.}\ \bibnamefont {Maldacena}}, \bibinfo
  {author} {\bibfnamefont {H.}~\bibnamefont {Ooguri}}, \ and\ \bibinfo {author}
  {\bibfnamefont {Y.}~\bibnamefont {Oz}},\ }\href {\doibase
  10.1016/S0370-1573(99)00083-6} {\bibfield  {journal} {\bibinfo  {journal}
  {Phys. Rept.}\ }\textbf {\bibinfo {volume} {323}},\ \bibinfo {pages} {183}
  (\bibinfo {year} {2000})},\ \Eprint {http://arxiv.org/abs/hep-th/9905111}
  {arXiv:hep-th/9905111 [hep-th]} \BibitemShut {NoStop}%
\bibitem [{\citenamefont {de~Haro}\ \emph {et~al.}(2001)\citenamefont
  {de~Haro}, \citenamefont {Solodukhin},\ and\ \citenamefont
  {Skenderis}}]{deHaro:2000vlm}%
  \BibitemOpen
  \bibfield  {author} {\bibinfo {author} {\bibfnamefont {S.}~\bibnamefont
  {de~Haro}}, \bibinfo {author} {\bibfnamefont {S.~N.}\ \bibnamefont
  {Solodukhin}}, \ and\ \bibinfo {author} {\bibfnamefont {K.}~\bibnamefont
  {Skenderis}},\ }\href {\doibase 10.1007/s002200100381} {\bibfield  {journal}
  {\bibinfo  {journal} {Commun. Math. Phys.}\ }\textbf {\bibinfo {volume}
  {217}},\ \bibinfo {pages} {595} (\bibinfo {year} {2001})},\ \Eprint
  {http://arxiv.org/abs/hep-th/0002230} {arXiv:hep-th/0002230} \BibitemShut
  {NoStop}%
\bibitem [{\citenamefont {Benna}\ \emph {et~al.}(2008)\citenamefont {Benna},
  \citenamefont {Klebanov}, \citenamefont {Klose},\ and\ \citenamefont
  {Smedback}}]{Benna:2008zy}%
  \BibitemOpen
  \bibfield  {author} {\bibinfo {author} {\bibfnamefont {M.}~\bibnamefont
  {Benna}}, \bibinfo {author} {\bibfnamefont {I.}~\bibnamefont {Klebanov}},
  \bibinfo {author} {\bibfnamefont {T.}~\bibnamefont {Klose}}, \ and\ \bibinfo
  {author} {\bibfnamefont {M.}~\bibnamefont {Smedback}},\ }\href {\doibase
  10.1088/1126-6708/2008/09/072} {\bibfield  {journal} {\bibinfo  {journal}
  {JHEP}\ }\textbf {\bibinfo {volume} {09}},\ \bibinfo {pages} {072} (\bibinfo
  {year} {2008})},\ \Eprint {http://arxiv.org/abs/0806.1519} {arXiv:0806.1519
  [hep-th]} \BibitemShut {NoStop}%
\bibitem [{\citenamefont {Bergman}\ \emph {et~al.}(2018)\citenamefont
  {Bergman}, \citenamefont {Rodr{\'\i}guez-G{\'o}mez},\ and\ \citenamefont
  {Uhlemann}}]{Bergman:2018hin}%
  \BibitemOpen
  \bibfield  {author} {\bibinfo {author} {\bibfnamefont {O.}~\bibnamefont
  {Bergman}}, \bibinfo {author} {\bibfnamefont {D.}~\bibnamefont
  {Rodr{\'\i}guez-G{\'o}mez}}, \ and\ \bibinfo {author} {\bibfnamefont {C.~F.}\
  \bibnamefont {Uhlemann}},\ }\href {\doibase 10.1007/JHEP08(2018)127}
  {\bibfield  {journal} {\bibinfo  {journal} {JHEP}\ }\textbf {\bibinfo
  {volume} {08}},\ \bibinfo {pages} {127} (\bibinfo {year} {2018})},\ \Eprint
  {http://arxiv.org/abs/1806.07898} {arXiv:1806.07898 [hep-th]} \BibitemShut
  {NoStop}%
\bibitem [{\citenamefont {Ryu}\ and\ \citenamefont
  {Takayanagi}(2006)}]{Ryu:2006ef}%
  \BibitemOpen
  \bibfield  {author} {\bibinfo {author} {\bibfnamefont {S.}~\bibnamefont
  {Ryu}}\ and\ \bibinfo {author} {\bibfnamefont {T.}~\bibnamefont
  {Takayanagi}},\ }\href {\doibase 10.1088/1126-6708/2006/08/045} {\bibfield
  {journal} {\bibinfo  {journal} {JHEP}\ }\textbf {\bibinfo {volume} {08}},\
  \bibinfo {pages} {045} (\bibinfo {year} {2006})},\ \Eprint
  {http://arxiv.org/abs/hep-th/0605073} {arXiv:hep-th/0605073 [hep-th]}
  \BibitemShut {NoStop}%
\bibitem [{\citenamefont {Swingle}(2012)}]{Swingle:2009bg}%
  \BibitemOpen
  \bibfield  {author} {\bibinfo {author} {\bibfnamefont {B.}~\bibnamefont
  {Swingle}},\ }\href {\doibase 10.1103/PhysRevD.86.065007} {\bibfield
  {journal} {\bibinfo  {journal} {Phys. Rev. D}\ }\textbf {\bibinfo {volume}
  {86}},\ \bibinfo {pages} {065007} (\bibinfo {year} {2012})},\ \Eprint
  {http://arxiv.org/abs/0905.1317} {arXiv:0905.1317 [cond-mat.str-el]}
  \BibitemShut {NoStop}%
\bibitem [{\citenamefont {Mojtahedi}\ \emph {et~al.}(2022)\citenamefont
  {Mojtahedi}, \citenamefont {Hamghalam}, \citenamefont {Do},\ and\
  \citenamefont {Simpson}}]{Mojtahedi_2022}%
  \BibitemOpen
  \bibfield  {author} {\bibinfo {author} {\bibfnamefont {R.}~\bibnamefont
  {Mojtahedi}}, \bibinfo {author} {\bibfnamefont {M.}~\bibnamefont
  {Hamghalam}}, \bibinfo {author} {\bibfnamefont {R.~K.~G.}\ \bibnamefont
  {Do}}, \ and\ \bibinfo {author} {\bibfnamefont {A.~L.}\ \bibnamefont
  {Simpson}},\ }in\ \href {\doibase 10.1007/978-3-031-18814-5_11} {\emph
  {\bibinfo {booktitle} {Multiscale Multimodal Medical Imaging}}}\ (\bibinfo
  {publisher} {Springer Nature Switzerland},\ \bibinfo {year} {2022})\ pp.\
  \bibinfo {pages} {110--120}\BibitemShut {NoStop}%
\bibitem [{\citenamefont {Park}\ \emph
  {et~al.}(2022{\natexlab{a}})\citenamefont {Park}, \citenamefont {Kim},\ and\
  \citenamefont {Lee}}]{Park:2022aa}%
  \BibitemOpen
  \bibfield  {author} {\bibinfo {author} {\bibfnamefont {C.}~\bibnamefont
  {Park}}, \bibinfo {author} {\bibfnamefont {S.-J.}\ \bibnamefont {Kim}}, \
  and\ \bibinfo {author} {\bibfnamefont {J.~H.}\ \bibnamefont {Lee}},\ }\href
  {https://arxiv.org/pdf/2212.01214.pdf} {\  (\bibinfo {year}
  {2022}{\natexlab{a}})},\ \Eprint {http://arxiv.org/abs/2212.01214}
  {2212.01214} \BibitemShut {NoStop}%
\bibitem [{\citenamefont {Park}\ \emph
  {et~al.}(2022{\natexlab{b}})\citenamefont {Park}, \citenamefont {Hwang},
  \citenamefont {Cho},\ and\ \citenamefont {Kim}}]{Park:2022fqy}%
  \BibitemOpen
  \bibfield  {author} {\bibinfo {author} {\bibfnamefont {C.}~\bibnamefont
  {Park}}, \bibinfo {author} {\bibfnamefont {C.-O.}\ \bibnamefont {Hwang}},
  \bibinfo {author} {\bibfnamefont {K.}~\bibnamefont {Cho}}, \ and\ \bibinfo
  {author} {\bibfnamefont {S.-J.}\ \bibnamefont {Kim}},\ }\href {\doibase
  10.1103/PhysRevD.106.106017} {\bibfield  {journal} {\bibinfo  {journal}
  {Phys. Rev. D}\ }\textbf {\bibinfo {volume} {106}},\ \bibinfo {pages}
  {106017} (\bibinfo {year} {2022}{\natexlab{b}})},\ \Eprint
  {http://arxiv.org/abs/2205.04445} {arXiv:2205.04445 [hep-th]} \BibitemShut
  {NoStop}%
\bibitem [{\citenamefont {Park}\ and\ \citenamefont {Lee}(2020)}]{Park:2020aa}%
  \BibitemOpen
  \bibfield  {author} {\bibinfo {author} {\bibfnamefont {C.}~\bibnamefont
  {Park}}\ and\ \bibinfo {author} {\bibfnamefont {J.~H.}\ \bibnamefont {Lee}},\
  }\href {\doibase 10.1103/PhysRevD.101.086008} {\bibfield  {journal} {\bibinfo
   {journal} {Phys. Rev. D}\ }\textbf {\bibinfo {volume} {101}},\ \bibinfo
  {pages} {086008} (\bibinfo {year} {2020})},\ \Eprint
  {http://arxiv.org/abs/1910.05741} {1910.05741} \BibitemShut {NoStop}%
\bibitem [{\citenamefont {Park}\ \emph {et~al.}(2018)\citenamefont {Park},
  \citenamefont {Ro},\ and\ \citenamefont {Lee}}]{Park:2018aa}%
  \BibitemOpen
  \bibfield  {author} {\bibinfo {author} {\bibfnamefont {C.}~\bibnamefont
  {Park}}, \bibinfo {author} {\bibfnamefont {D.}~\bibnamefont {Ro}}, \ and\
  \bibinfo {author} {\bibfnamefont {J.~H.}\ \bibnamefont {Lee}},\ }\href
  {\doibase 10.1007/JHEP11(2018)165} {\  (\bibinfo {year} {2018}),\
  10.1007/JHEP11(2018)165},\ \Eprint {http://arxiv.org/abs/1806.09072}
  {1806.09072} \BibitemShut {NoStop}%
\bibitem [{\citenamefont {Erlich}\ \emph {et~al.}(2005)\citenamefont {Erlich},
  \citenamefont {Katz}, \citenamefont {Son},\ and\ \citenamefont
  {Stephanov}}]{Erlich:2005qh}%
  \BibitemOpen
  \bibfield  {author} {\bibinfo {author} {\bibfnamefont {J.}~\bibnamefont
  {Erlich}}, \bibinfo {author} {\bibfnamefont {E.}~\bibnamefont {Katz}},
  \bibinfo {author} {\bibfnamefont {D.~T.}\ \bibnamefont {Son}}, \ and\
  \bibinfo {author} {\bibfnamefont {M.~A.}\ \bibnamefont {Stephanov}},\ }\href
  {\doibase 10.1103/PhysRevLett.95.261602} {\bibfield  {journal} {\bibinfo
  {journal} {Phys. Rev. Lett.}\ }\textbf {\bibinfo {volume} {95}},\ \bibinfo
  {pages} {261602} (\bibinfo {year} {2005})},\ \Eprint
  {http://arxiv.org/abs/hep-ph/0501128} {arXiv:hep-ph/0501128} \BibitemShut
  {NoStop}%
\bibitem [{\citenamefont {Karch}\ \emph {et~al.}(2006)\citenamefont {Karch},
  \citenamefont {Katz}, \citenamefont {Son},\ and\ \citenamefont
  {Stephanov}}]{Karch:2006pv}%
  \BibitemOpen
  \bibfield  {author} {\bibinfo {author} {\bibfnamefont {A.}~\bibnamefont
  {Karch}}, \bibinfo {author} {\bibfnamefont {E.}~\bibnamefont {Katz}},
  \bibinfo {author} {\bibfnamefont {D.~T.}\ \bibnamefont {Son}}, \ and\
  \bibinfo {author} {\bibfnamefont {M.~A.}\ \bibnamefont {Stephanov}},\ }\href
  {\doibase 10.1103/PhysRevD.74.015005} {\bibfield  {journal} {\bibinfo
  {journal} {Phys. Rev. D}\ }\textbf {\bibinfo {volume} {74}},\ \bibinfo
  {pages} {015005} (\bibinfo {year} {2006})},\ \Eprint
  {http://arxiv.org/abs/hep-ph/0602229} {arXiv:hep-ph/0602229} \BibitemShut
  {NoStop}%
\bibitem [{\citenamefont {Karch}\ and\ \citenamefont
  {Katz}(2002)}]{Karch:2002sh}%
  \BibitemOpen
  \bibfield  {author} {\bibinfo {author} {\bibfnamefont {A.}~\bibnamefont
  {Karch}}\ and\ \bibinfo {author} {\bibfnamefont {E.}~\bibnamefont {Katz}},\
  }\href {\doibase 10.1088/1126-6708/2002/06/043} {\bibfield  {journal}
  {\bibinfo  {journal} {JHEP}\ }\textbf {\bibinfo {volume} {06}},\ \bibinfo
  {pages} {043} (\bibinfo {year} {2002})},\ \Eprint
  {http://arxiv.org/abs/hep-th/0205236} {arXiv:hep-th/0205236} \BibitemShut
  {NoStop}%
\bibitem [{\citenamefont {Casini}\ \emph {et~al.}(2011)\citenamefont {Casini},
  \citenamefont {Huerta},\ and\ \citenamefont {Myers}}]{Casini:2011kv}%
  \BibitemOpen
  \bibfield  {author} {\bibinfo {author} {\bibfnamefont {H.}~\bibnamefont
  {Casini}}, \bibinfo {author} {\bibfnamefont {M.}~\bibnamefont {Huerta}}, \
  and\ \bibinfo {author} {\bibfnamefont {R.~C.}\ \bibnamefont {Myers}},\ }\href
  {\doibase 10.1007/JHEP05(2011)036} {\bibfield  {journal} {\bibinfo  {journal}
  {JHEP}\ }\textbf {\bibinfo {volume} {05}},\ \bibinfo {pages} {036} (\bibinfo
  {year} {2011})},\ \Eprint {http://arxiv.org/abs/1102.0440} {arXiv:1102.0440
  [hep-th]} \BibitemShut {NoStop}%
\bibitem [{\citenamefont {Faulkner}\ \emph {et~al.}(2013)\citenamefont
  {Faulkner}, \citenamefont {Lewkowycz},\ and\ \citenamefont
  {Maldacena}}]{Faulkner:2013ana}%
  \BibitemOpen
  \bibfield  {author} {\bibinfo {author} {\bibfnamefont {T.}~\bibnamefont
  {Faulkner}}, \bibinfo {author} {\bibfnamefont {A.}~\bibnamefont {Lewkowycz}},
  \ and\ \bibinfo {author} {\bibfnamefont {J.}~\bibnamefont {Maldacena}},\
  }\href {\doibase 10.1007/JHEP11(2013)074} {\bibfield  {journal} {\bibinfo
  {journal} {JHEP}\ }\textbf {\bibinfo {volume} {11}},\ \bibinfo {pages} {074}
  (\bibinfo {year} {2013})},\ \Eprint {http://arxiv.org/abs/1307.2892}
  {arXiv:1307.2892 [hep-th]} \BibitemShut {NoStop}%
\bibitem [{\citenamefont {Myers}\ and\ \citenamefont
  {Sinha}(2011)}]{Myers:2010tj}%
  \BibitemOpen
  \bibfield  {author} {\bibinfo {author} {\bibfnamefont {R.~C.}\ \bibnamefont
  {Myers}}\ and\ \bibinfo {author} {\bibfnamefont {A.}~\bibnamefont {Sinha}},\
  }\href {\doibase 10.1007/JHEP01(2011)125} {\bibfield  {journal} {\bibinfo
  {journal} {JHEP}\ }\textbf {\bibinfo {volume} {01}},\ \bibinfo {pages} {125}
  (\bibinfo {year} {2011})},\ \Eprint {http://arxiv.org/abs/1011.5819}
  {arXiv:1011.5819 [hep-th]} \BibitemShut {NoStop}%
\bibitem [{\citenamefont {Maldacena}\ and\ \citenamefont
  {Susskind}(2013)}]{Maldacena:2013xja}%
  \BibitemOpen
  \bibfield  {author} {\bibinfo {author} {\bibfnamefont {J.}~\bibnamefont
  {Maldacena}}\ and\ \bibinfo {author} {\bibfnamefont {L.}~\bibnamefont
  {Susskind}},\ }\href {\doibase 10.1002/prop.201300020} {\bibfield  {journal}
  {\bibinfo  {journal} {Fortsch. Phys.}\ }\textbf {\bibinfo {volume} {61}},\
  \bibinfo {pages} {781} (\bibinfo {year} {2013})},\ \Eprint
  {http://arxiv.org/abs/1306.0533} {arXiv:1306.0533 [hep-th]} \BibitemShut
  {NoStop}%
\bibitem [{\citenamefont {Lewkowycz}\ and\ \citenamefont
  {Maldacena}(2013)}]{Lewkowycz:2013nqa}%
  \BibitemOpen
  \bibfield  {author} {\bibinfo {author} {\bibfnamefont {A.}~\bibnamefont
  {Lewkowycz}}\ and\ \bibinfo {author} {\bibfnamefont {J.}~\bibnamefont
  {Maldacena}},\ }\href {\doibase 10.1007/JHEP08(2013)090} {\bibfield
  {journal} {\bibinfo  {journal} {JHEP}\ }\textbf {\bibinfo {volume} {08}},\
  \bibinfo {pages} {090} (\bibinfo {year} {2013})},\ \Eprint
  {http://arxiv.org/abs/1304.4926} {arXiv:1304.4926 [hep-th]} \BibitemShut
  {NoStop}%
\bibitem [{\citenamefont {Nishioka}\ \emph {et~al.}(2009)\citenamefont
  {Nishioka}, \citenamefont {Ryu},\ and\ \citenamefont
  {Takayanagi}}]{Nishioka:2009un}%
  \BibitemOpen
  \bibfield  {author} {\bibinfo {author} {\bibfnamefont {T.}~\bibnamefont
  {Nishioka}}, \bibinfo {author} {\bibfnamefont {S.}~\bibnamefont {Ryu}}, \
  and\ \bibinfo {author} {\bibfnamefont {T.}~\bibnamefont {Takayanagi}},\
  }\href {\doibase 10.1088/1751-8113/42/50/504008} {\bibfield  {journal}
  {\bibinfo  {journal} {J. Phys. A}\ }\textbf {\bibinfo {volume} {42}},\
  \bibinfo {pages} {504008} (\bibinfo {year} {2009})},\ \Eprint
  {http://arxiv.org/abs/0905.0932} {arXiv:0905.0932 [hep-th]} \BibitemShut
  {NoStop}%
\bibitem [{\citenamefont {Penington}(2020)}]{Penington:2019npb}%
  \BibitemOpen
  \bibfield  {author} {\bibinfo {author} {\bibfnamefont {G.}~\bibnamefont
  {Penington}},\ }\href {\doibase 10.1007/JHEP09(2020)002} {\bibfield
  {journal} {\bibinfo  {journal} {JHEP}\ }\textbf {\bibinfo {volume} {09}},\
  \bibinfo {pages} {002} (\bibinfo {year} {2020})},\ \Eprint
  {http://arxiv.org/abs/1905.08255} {arXiv:1905.08255 [hep-th]} \BibitemShut
  {NoStop}%
\bibitem [{\citenamefont {You}\ \emph {et~al.}(2018)\citenamefont {You},
  \citenamefont {Yang},\ and\ \citenamefont {Qi}}]{You_2018}%
  \BibitemOpen
  \bibfield  {author} {\bibinfo {author} {\bibfnamefont {Y.-Z.}\ \bibnamefont
  {You}}, \bibinfo {author} {\bibfnamefont {Z.}~\bibnamefont {Yang}}, \ and\
  \bibinfo {author} {\bibfnamefont {X.-L.}\ \bibnamefont {Qi}},\ }\href
  {\doibase 10.1103/physrevb.97.045153} {\bibfield  {journal} {\bibinfo
  {journal} {Physical Review B}\ }\textbf {\bibinfo {volume} {97}} (\bibinfo
  {year} {2018}),\ 10.1103/physrevb.97.045153}\BibitemShut {NoStop}%
\bibitem [{\citenamefont {Hashimoto}\ \emph {et~al.}(2018)\citenamefont
  {Hashimoto}, \citenamefont {Sugishita}, \citenamefont {Tanaka},\ and\
  \citenamefont {Tomiya}}]{Hashimoto:2018wm}%
  \BibitemOpen
  \bibfield  {author} {\bibinfo {author} {\bibfnamefont {K.}~\bibnamefont
  {Hashimoto}}, \bibinfo {author} {\bibfnamefont {S.}~\bibnamefont
  {Sugishita}}, \bibinfo {author} {\bibfnamefont {A.}~\bibnamefont {Tanaka}}, \
  and\ \bibinfo {author} {\bibfnamefont {A.}~\bibnamefont {Tomiya}},\ }\href
  {\doibase 10.1103/PhysRevD.98.046019} {\bibfield  {journal} {\bibinfo
  {journal} {Phys. Rev. D}\ }\textbf {\bibinfo {volume} {98}},\ \bibinfo
  {pages} {046019} (\bibinfo {year} {2018})},\ \Eprint
  {http://arxiv.org/abs/1802.08313} {1802.08313} \BibitemShut {NoStop}%
\bibitem [{\citenamefont {Shiba~Funai}\ and\ \citenamefont
  {Giataganas}(2020)}]{ShibaFunai:2018aaw}%
  \BibitemOpen
  \bibfield  {author} {\bibinfo {author} {\bibfnamefont {S.}~\bibnamefont
  {Shiba~Funai}}\ and\ \bibinfo {author} {\bibfnamefont {D.}~\bibnamefont
  {Giataganas}},\ }\href {\doibase 10.1103/PhysRevResearch.2.033415} {\bibfield
   {journal} {\bibinfo  {journal} {Phys. Rev. Res.}\ }\textbf {\bibinfo
  {volume} {2}},\ \bibinfo {pages} {033415} (\bibinfo {year} {2020})},\ \Eprint
  {http://arxiv.org/abs/1810.08179} {arXiv:1810.08179 [cond-mat.stat-mech]}
  \BibitemShut {NoStop}%
\bibitem [{\citenamefont {Akutagawa}\ \emph {et~al.}(2020)\citenamefont
  {Akutagawa}, \citenamefont {Hashimoto},\ and\ \citenamefont
  {Sumimoto}}]{Akutagawa:2020aa}%
  \BibitemOpen
  \bibfield  {author} {\bibinfo {author} {\bibfnamefont {T.}~\bibnamefont
  {Akutagawa}}, \bibinfo {author} {\bibfnamefont {K.}~\bibnamefont
  {Hashimoto}}, \ and\ \bibinfo {author} {\bibfnamefont {T.}~\bibnamefont
  {Sumimoto}},\ }\href {\doibase 10.1103/PhysRevD.102.026020} {\bibfield
  {journal} {\bibinfo  {journal} {Phys. Rev. D}\ }\textbf {\bibinfo {volume}
  {102}},\ \bibinfo {pages} {026020} (\bibinfo {year} {2020})},\ \Eprint
  {http://arxiv.org/abs/2005.02636} {2005.02636} \BibitemShut {NoStop}%
\bibitem [{\citenamefont {Chen}\ \emph {et~al.}(2021)\citenamefont {Chen},
  \citenamefont {Ding}, \citenamefont {Liu}, \citenamefont {Papp},\ and\
  \citenamefont {Yang}}]{Chen:2021giw}%
  \BibitemOpen
  \bibfield  {author} {\bibinfo {author} {\bibfnamefont {S.~Y.}\ \bibnamefont
  {Chen}}, \bibinfo {author} {\bibfnamefont {H.~T.}\ \bibnamefont {Ding}},
  \bibinfo {author} {\bibfnamefont {F.~Y.}\ \bibnamefont {Liu}}, \bibinfo
  {author} {\bibfnamefont {G.}~\bibnamefont {Papp}}, \ and\ \bibinfo {author}
  {\bibfnamefont {C.~B.}\ \bibnamefont {Yang}},\ }\href@noop {} {\  (\bibinfo
  {year} {2021})},\ \Eprint {http://arxiv.org/abs/2110.13521} {arXiv:2110.13521
  [hep-lat]} \BibitemShut {NoStop}%
\bibitem [{\citenamefont {Lam}\ and\ \citenamefont {You}(2021)}]{Lam:2021ugb}%
  \BibitemOpen
  \bibfield  {author} {\bibinfo {author} {\bibfnamefont {J.}~\bibnamefont
  {Lam}}\ and\ \bibinfo {author} {\bibfnamefont {Y.-Z.}\ \bibnamefont {You}},\
  }\href {\doibase 10.1103/PhysRevResearch.3.043199} {\bibfield  {journal}
  {\bibinfo  {journal} {Phys. Rev. Res.}\ }\textbf {\bibinfo {volume} {3}},\
  \bibinfo {pages} {043199} (\bibinfo {year} {2021})},\ \Eprint
  {http://arxiv.org/abs/2110.01115} {arXiv:2110.01115 [hep-th]} \BibitemShut
  {NoStop}%
\bibitem [{\citenamefont {Li}\ \emph {et~al.}(2023)\citenamefont {Li},
  \citenamefont {Ling}, \citenamefont {Liu},\ and\ \citenamefont
  {Wu}}]{Li:2023aa}%
  \BibitemOpen
  \bibfield  {author} {\bibinfo {author} {\bibfnamefont {K.}~\bibnamefont
  {Li}}, \bibinfo {author} {\bibfnamefont {Y.}~\bibnamefont {Ling}}, \bibinfo
  {author} {\bibfnamefont {P.}~\bibnamefont {Liu}}, \ and\ \bibinfo {author}
  {\bibfnamefont {M.-H.}\ \bibnamefont {Wu}},\ }\href {\doibase
  10.1103/PhysRevD.107.066021} {\bibfield  {journal} {\bibinfo  {journal}
  {Phys. Rev. D 107,}\ }\textbf {\bibinfo {volume} {no.6}},\ \bibinfo {pages}
  {066021} (\bibinfo {year} {2023})},\ \Eprint
  {http://arxiv.org/abs/2209.05203} {2209.05203} \BibitemShut {NoStop}%
\bibitem [{\citenamefont {Jokela}\ and\ \citenamefont
  {P{\"o}nni}(2021)}]{Jokela:2021aa}%
  \BibitemOpen
  \bibfield  {author} {\bibinfo {author} {\bibfnamefont {N.}~\bibnamefont
  {Jokela}}\ and\ \bibinfo {author} {\bibfnamefont {A.}~\bibnamefont
  {P{\"o}nni}},\ }\href {\doibase 10.1103/PhysRevD.103.026010} {\bibfield
  {journal} {\bibinfo  {journal} {Phys. Rev. D}\ }\textbf {\bibinfo {volume}
  {103}},\ \bibinfo {pages} {026010} (\bibinfo {year} {2021})},\ \Eprint
  {http://arxiv.org/abs/2007.00010} {2007.00010} \BibitemShut {NoStop}%
\bibitem [{\citenamefont {Jokela}\ \emph {et~al.}(2023)\citenamefont {Jokela},
  \citenamefont {P{\"o}nni}, \citenamefont {Rindlisbacher}, \citenamefont
  {Rummukainen},\ and\ \citenamefont {Salami}}]{Jokela:2023aa}%
  \BibitemOpen
  \bibfield  {author} {\bibinfo {author} {\bibfnamefont {N.}~\bibnamefont
  {Jokela}}, \bibinfo {author} {\bibfnamefont {A.}~\bibnamefont {P{\"o}nni}},
  \bibinfo {author} {\bibfnamefont {T.}~\bibnamefont {Rindlisbacher}}, \bibinfo
  {author} {\bibfnamefont {K.}~\bibnamefont {Rummukainen}}, \ and\ \bibinfo
  {author} {\bibfnamefont {A.}~\bibnamefont {Salami}},\ }\href {\doibase
  10.1007/JHEP12(2023)137} {\bibfield  {journal} {\bibinfo  {journal} {J. High
  Energ. Phys.}\ }\textbf {\bibinfo {volume} {2023}},\ \bibinfo {pages} {137}
  (\bibinfo {year} {2023})},\ \Eprint {http://arxiv.org/abs/2304.08949}
  {2304.08949} \BibitemShut {NoStop}%
\bibitem [{\citenamefont {Park}\ and\ \citenamefont
  {Lee}(2024)}]{Park:2024fik}%
  \BibitemOpen
  \bibfield  {author} {\bibinfo {author} {\bibfnamefont {C.}~\bibnamefont
  {Park}}\ and\ \bibinfo {author} {\bibfnamefont {J.~H.}\ \bibnamefont {Lee}},\
  }\href@noop {} {\  (\bibinfo {year} {2024})},\ \Eprint
  {http://arxiv.org/abs/2406.17221} {arXiv:2406.17221 [hep-th]} \BibitemShut
  {NoStop}%
\bibitem [{\citenamefont {Vaswani}\ \emph {et~al.}(2017)\citenamefont
  {Vaswani}, \citenamefont {Shazeer}, \citenamefont {Parmar}, \citenamefont
  {Uszkoreit}, \citenamefont {Jones}, \citenamefont {Gomez}, \citenamefont
  {Kaiser},\ and\ \citenamefont {Polosukhin}}]{Vaswani:2017aa}%
  \BibitemOpen
  \bibfield  {author} {\bibinfo {author} {\bibfnamefont {A.}~\bibnamefont
  {Vaswani}}, \bibinfo {author} {\bibfnamefont {N.}~\bibnamefont {Shazeer}},
  \bibinfo {author} {\bibfnamefont {N.}~\bibnamefont {Parmar}}, \bibinfo
  {author} {\bibfnamefont {J.}~\bibnamefont {Uszkoreit}}, \bibinfo {author}
  {\bibfnamefont {L.}~\bibnamefont {Jones}}, \bibinfo {author} {\bibfnamefont
  {A.~N.}\ \bibnamefont {Gomez}}, \bibinfo {author} {\bibfnamefont
  {L.}~\bibnamefont {Kaiser}}, \ and\ \bibinfo {author} {\bibfnamefont
  {I.}~\bibnamefont {Polosukhin}},\ }\href
  {https://arxiv.org/pdf/1706.03762.pdf} {\  (\bibinfo {year} {2017})},\
  \Eprint {http://arxiv.org/abs/1706.03762} {1706.03762} \BibitemShut {NoStop}%
\bibitem [{\citenamefont {Brown}\ \emph {et~al.}(2020)\citenamefont {Brown},
  \citenamefont {Mann}, \citenamefont {Ryder}, \citenamefont {Subbiah},
  \citenamefont {Kaplan}, \citenamefont {Dhariwal}, \citenamefont
  {Neelakantan}, \citenamefont {Shyam}, \citenamefont {Sastry}, \citenamefont
  {Askell}, \citenamefont {Agarwal}, \citenamefont {Herbert-Voss},
  \citenamefont {Krueger}, \citenamefont {Henighan}, \citenamefont {Child},
  \citenamefont {Ramesh}, \citenamefont {Ziegler}, \citenamefont {Wu},
  \citenamefont {Winter}, \citenamefont {Hesse}, \citenamefont {Chen},
  \citenamefont {Sigler}, \citenamefont {Litwin}, \citenamefont {Gray},
  \citenamefont {Chess}, \citenamefont {Clark}, \citenamefont {Berner},
  \citenamefont {McCandlish}, \citenamefont {Radford}, \citenamefont
  {Sutskever},\ and\ \citenamefont {Amodei}}]{Brown:2020aa}%
  \BibitemOpen
  \bibfield  {author} {\bibinfo {author} {\bibfnamefont {T.~B.}\ \bibnamefont
  {Brown}}, \bibinfo {author} {\bibfnamefont {B.}~\bibnamefont {Mann}},
  \bibinfo {author} {\bibfnamefont {N.}~\bibnamefont {Ryder}}, \bibinfo
  {author} {\bibfnamefont {M.}~\bibnamefont {Subbiah}}, \bibinfo {author}
  {\bibfnamefont {J.}~\bibnamefont {Kaplan}}, \bibinfo {author} {\bibfnamefont
  {P.}~\bibnamefont {Dhariwal}}, \bibinfo {author} {\bibfnamefont
  {A.}~\bibnamefont {Neelakantan}}, \bibinfo {author} {\bibfnamefont
  {P.}~\bibnamefont {Shyam}}, \bibinfo {author} {\bibfnamefont
  {G.}~\bibnamefont {Sastry}}, \bibinfo {author} {\bibfnamefont
  {A.}~\bibnamefont {Askell}}, \bibinfo {author} {\bibfnamefont
  {S.}~\bibnamefont {Agarwal}}, \bibinfo {author} {\bibfnamefont
  {A.}~\bibnamefont {Herbert-Voss}}, \bibinfo {author} {\bibfnamefont
  {G.}~\bibnamefont {Krueger}}, \bibinfo {author} {\bibfnamefont
  {T.}~\bibnamefont {Henighan}}, \bibinfo {author} {\bibfnamefont
  {R.}~\bibnamefont {Child}}, \bibinfo {author} {\bibfnamefont
  {A.}~\bibnamefont {Ramesh}}, \bibinfo {author} {\bibfnamefont {D.~M.}\
  \bibnamefont {Ziegler}}, \bibinfo {author} {\bibfnamefont {J.}~\bibnamefont
  {Wu}}, \bibinfo {author} {\bibfnamefont {C.}~\bibnamefont {Winter}}, \bibinfo
  {author} {\bibfnamefont {C.}~\bibnamefont {Hesse}}, \bibinfo {author}
  {\bibfnamefont {M.}~\bibnamefont {Chen}}, \bibinfo {author} {\bibfnamefont
  {E.}~\bibnamefont {Sigler}}, \bibinfo {author} {\bibfnamefont
  {M.}~\bibnamefont {Litwin}}, \bibinfo {author} {\bibfnamefont
  {S.}~\bibnamefont {Gray}}, \bibinfo {author} {\bibfnamefont {B.}~\bibnamefont
  {Chess}}, \bibinfo {author} {\bibfnamefont {J.}~\bibnamefont {Clark}},
  \bibinfo {author} {\bibfnamefont {C.}~\bibnamefont {Berner}}, \bibinfo
  {author} {\bibfnamefont {S.}~\bibnamefont {McCandlish}}, \bibinfo {author}
  {\bibfnamefont {A.}~\bibnamefont {Radford}}, \bibinfo {author} {\bibfnamefont
  {I.}~\bibnamefont {Sutskever}}, \ and\ \bibinfo {author} {\bibfnamefont
  {D.}~\bibnamefont {Amodei}},\ }\href {https://arxiv.org/pdf/2005.14165.pdf}
  {\  (\bibinfo {year} {2020})},\ \Eprint {http://arxiv.org/abs/2005.14165}
  {2005.14165} \BibitemShut {NoStop}%
\bibitem [{\citenamefont {Devlin}\ \emph {et~al.}(2018)\citenamefont {Devlin},
  \citenamefont {Chang}, \citenamefont {Lee},\ and\ \citenamefont
  {Toutanova}}]{Devlin:2018aa}%
  \BibitemOpen
  \bibfield  {author} {\bibinfo {author} {\bibfnamefont {J.}~\bibnamefont
  {Devlin}}, \bibinfo {author} {\bibfnamefont {M.-W.}\ \bibnamefont {Chang}},
  \bibinfo {author} {\bibfnamefont {K.}~\bibnamefont {Lee}}, \ and\ \bibinfo
  {author} {\bibfnamefont {K.}~\bibnamefont {Toutanova}},\ }\href
  {https://arxiv.org/pdf/1810.04805.pdf} {\  (\bibinfo {year} {2018})},\
  \Eprint {http://arxiv.org/abs/1810.04805} {1810.04805} \BibitemShut {NoStop}%
\bibitem [{\citenamefont {Touvron}\ \emph {et~al.}(2023)\citenamefont
  {Touvron}, \citenamefont {Lavril}, \citenamefont {Izacard}, \citenamefont
  {Martinet}, \citenamefont {Lachaux}, \citenamefont {Lacroix}, \citenamefont
  {Rozi{\`e}re}, \citenamefont {Goyal}, \citenamefont {Hambro}, \citenamefont
  {Azhar}, \citenamefont {Rodriguez}, \citenamefont {Joulin}, \citenamefont
  {Grave},\ and\ \citenamefont {Lample}}]{Touvron:2023aa}%
  \BibitemOpen
  \bibfield  {author} {\bibinfo {author} {\bibfnamefont {H.}~\bibnamefont
  {Touvron}}, \bibinfo {author} {\bibfnamefont {T.}~\bibnamefont {Lavril}},
  \bibinfo {author} {\bibfnamefont {G.}~\bibnamefont {Izacard}}, \bibinfo
  {author} {\bibfnamefont {X.}~\bibnamefont {Martinet}}, \bibinfo {author}
  {\bibfnamefont {M.-A.}\ \bibnamefont {Lachaux}}, \bibinfo {author}
  {\bibfnamefont {T.}~\bibnamefont {Lacroix}}, \bibinfo {author} {\bibfnamefont
  {B.}~\bibnamefont {Rozi{\`e}re}}, \bibinfo {author} {\bibfnamefont
  {N.}~\bibnamefont {Goyal}}, \bibinfo {author} {\bibfnamefont
  {E.}~\bibnamefont {Hambro}}, \bibinfo {author} {\bibfnamefont
  {F.}~\bibnamefont {Azhar}}, \bibinfo {author} {\bibfnamefont
  {A.}~\bibnamefont {Rodriguez}}, \bibinfo {author} {\bibfnamefont
  {A.}~\bibnamefont {Joulin}}, \bibinfo {author} {\bibfnamefont
  {E.}~\bibnamefont {Grave}}, \ and\ \bibinfo {author} {\bibfnamefont
  {G.}~\bibnamefont {Lample}},\ }\href {https://arxiv.org/pdf/2302.13971.pdf}
  {\  (\bibinfo {year} {2023})},\ \Eprint {http://arxiv.org/abs/2302.13971}
  {2302.13971} \BibitemShut {NoStop}%
\bibitem [{\citenamefont {OpenAI}(2023)}]{openai2023gpt4}%
  \BibitemOpen
  \bibfield  {author} {\bibinfo {author} {\bibnamefont {OpenAI}},\ }\href@noop
  {} {\enquote {\bibinfo {title} {Gpt-4 technical report},}\ } (\bibinfo {year}
  {2023}),\ \Eprint {http://arxiv.org/abs/2303.08774} {arXiv:2303.08774
  [cs.CL]} \BibitemShut {NoStop}%
\bibitem [{\citenamefont {Dosovitskiy}\ \emph {et~al.}(2020)\citenamefont
  {Dosovitskiy}, \citenamefont {Beyer}, \citenamefont {Kolesnikov},
  \citenamefont {Weissenborn}, \citenamefont {Zhai}, \citenamefont
  {Unterthiner}, \citenamefont {Dehghani}, \citenamefont {Minderer},
  \citenamefont {Heigold}, \citenamefont {Gelly}, \citenamefont {Uszkoreit},\
  and\ \citenamefont {Houlsby}}]{Dosovitskiy:2020aa}%
  \BibitemOpen
  \bibfield  {author} {\bibinfo {author} {\bibfnamefont {A.}~\bibnamefont
  {Dosovitskiy}}, \bibinfo {author} {\bibfnamefont {L.}~\bibnamefont {Beyer}},
  \bibinfo {author} {\bibfnamefont {A.}~\bibnamefont {Kolesnikov}}, \bibinfo
  {author} {\bibfnamefont {D.}~\bibnamefont {Weissenborn}}, \bibinfo {author}
  {\bibfnamefont {X.}~\bibnamefont {Zhai}}, \bibinfo {author} {\bibfnamefont
  {T.}~\bibnamefont {Unterthiner}}, \bibinfo {author} {\bibfnamefont
  {M.}~\bibnamefont {Dehghani}}, \bibinfo {author} {\bibfnamefont
  {M.}~\bibnamefont {Minderer}}, \bibinfo {author} {\bibfnamefont
  {G.}~\bibnamefont {Heigold}}, \bibinfo {author} {\bibfnamefont
  {S.}~\bibnamefont {Gelly}}, \bibinfo {author} {\bibfnamefont
  {J.}~\bibnamefont {Uszkoreit}}, \ and\ \bibinfo {author} {\bibfnamefont
  {N.}~\bibnamefont {Houlsby}},\ }\href {https://arxiv.org/pdf/2010.11929.pdf}
  {\  (\bibinfo {year} {2020})},\ \Eprint {http://arxiv.org/abs/2010.11929}
  {2010.11929} \BibitemShut {NoStop}%
\bibitem [{\citenamefont {Qin}\ \emph {et~al.}(2023)\citenamefont {Qin},
  \citenamefont {Wang}, \citenamefont {Deng}, \citenamefont {Wang},
  \citenamefont {Chen}, \citenamefont {Hu},\ and\ \citenamefont
  {Deng}}]{Qin:2023aa}%
  \BibitemOpen
  \bibfield  {author} {\bibinfo {author} {\bibfnamefont {L.}~\bibnamefont
  {Qin}}, \bibinfo {author} {\bibfnamefont {M.}~\bibnamefont {Wang}}, \bibinfo
  {author} {\bibfnamefont {C.}~\bibnamefont {Deng}}, \bibinfo {author}
  {\bibfnamefont {K.}~\bibnamefont {Wang}}, \bibinfo {author} {\bibfnamefont
  {X.}~\bibnamefont {Chen}}, \bibinfo {author} {\bibfnamefont {J.}~\bibnamefont
  {Hu}}, \ and\ \bibinfo {author} {\bibfnamefont {W.}~\bibnamefont {Deng}},\
  }\href {\doibase https://doi.org/10.1109/TCSVT.2023.3304724} {\  (\bibinfo
  {year} {2023}),\ https://doi.org/10.1109/TCSVT.2023.3304724},\ \Eprint
  {http://arxiv.org/abs/2308.11509} {2308.11509} \BibitemShut {NoStop}%
\bibitem [{\citenamefont {Bi}\ \emph {et~al.}(2023)\citenamefont {Bi},
  \citenamefont {Xie}, \citenamefont {Zhang}, \citenamefont {Chen},
  \citenamefont {Gu},\ and\ \citenamefont {Tian}}]{Bi:2023aa}%
  \BibitemOpen
  \bibfield  {author} {\bibinfo {author} {\bibfnamefont {K.}~\bibnamefont
  {Bi}}, \bibinfo {author} {\bibfnamefont {L.}~\bibnamefont {Xie}}, \bibinfo
  {author} {\bibfnamefont {H.}~\bibnamefont {Zhang}}, \bibinfo {author}
  {\bibfnamefont {X.}~\bibnamefont {Chen}}, \bibinfo {author} {\bibfnamefont
  {X.}~\bibnamefont {Gu}}, \ and\ \bibinfo {author} {\bibfnamefont
  {Q.}~\bibnamefont {Tian}},\ }\href {\doibase 10.1038/s41586-023-06185-3}
  {\bibfield  {journal} {\bibinfo  {journal} {Nature}\ }\textbf {\bibinfo
  {volume} {619}},\ \bibinfo {pages} {533} (\bibinfo {year}
  {2023})}\BibitemShut {NoStop}%
\bibitem [{\citenamefont {Popel}\ and\ \citenamefont
  {Bojar}(2018)}]{Popel_2018}%
  \BibitemOpen
  \bibfield  {author} {\bibinfo {author} {\bibfnamefont {M.}~\bibnamefont
  {Popel}}\ and\ \bibinfo {author} {\bibfnamefont {O.}~\bibnamefont {Bojar}},\
  }\href {\doibase 10.2478/pralin-2018-0002} {\bibfield  {journal} {\bibinfo
  {journal} {The Prague Bulletin of Mathematical Linguistics}\ }\textbf
  {\bibinfo {volume} {110}},\ \bibinfo {pages} {43} (\bibinfo {year}
  {2018})}\BibitemShut {NoStop}%
\bibitem [{\citenamefont {Lee}\ \emph {et~al.}(2009)\citenamefont {Lee},
  \citenamefont {Park},\ and\ \citenamefont {Sin}}]{Lee:2009bya}%
  \BibitemOpen
  \bibfield  {author} {\bibinfo {author} {\bibfnamefont {B.-H.}\ \bibnamefont
  {Lee}}, \bibinfo {author} {\bibfnamefont {C.}~\bibnamefont {Park}}, \ and\
  \bibinfo {author} {\bibfnamefont {S.-J.}\ \bibnamefont {Sin}},\ }\href
  {\doibase 10.1088/1126-6708/2009/07/087} {\bibfield  {journal} {\bibinfo
  {journal} {JHEP}\ }\textbf {\bibinfo {volume} {07}},\ \bibinfo {pages} {087}
  (\bibinfo {year} {2009})},\ \Eprint {http://arxiv.org/abs/0905.2800}
  {arXiv:0905.2800 [hep-th]} \BibitemShut {NoStop}%
\bibitem [{\citenamefont {Park}(2010)}]{Park:2009nb}%
  \BibitemOpen
  \bibfield  {author} {\bibinfo {author} {\bibfnamefont {C.}~\bibnamefont
  {Park}},\ }\href {\doibase 10.1103/PhysRevD.81.045009} {\bibfield  {journal}
  {\bibinfo  {journal} {Phys. Rev. D}\ }\textbf {\bibinfo {volume} {81}},\
  \bibinfo {pages} {045009} (\bibinfo {year} {2010})},\ \Eprint
  {http://arxiv.org/abs/0907.0064} {arXiv:0907.0064 [hep-ph]} \BibitemShut
  {NoStop}%
\bibitem [{\citenamefont {Park}\ \emph {et~al.}(2024)\citenamefont {Park},
  \citenamefont {Kim},\ and\ \citenamefont {Lee}}]{Park:2022abi}%
  \BibitemOpen
  \bibfield  {author} {\bibinfo {author} {\bibfnamefont {C.}~\bibnamefont
  {Park}}, \bibinfo {author} {\bibfnamefont {S.-J.}\ \bibnamefont {Kim}}, \
  and\ \bibinfo {author} {\bibfnamefont {J.~H.}\ \bibnamefont {Lee}},\ }\href
  {\doibase 10.1142/S0217732324500044} {\bibfield  {journal} {\bibinfo
  {journal} {Mod. Phys. Lett. A}\ }\textbf {\bibinfo {volume} {39}},\ \bibinfo
  {pages} {2450004} (\bibinfo {year} {2024})},\ \Eprint
  {http://arxiv.org/abs/2212.01214} {arXiv:2212.01214 [hep-th]} \BibitemShut
  {NoStop}%
\bibitem [{\citenamefont {Park}(2023)}]{Park:2021wep}%
  \BibitemOpen
  \bibfield  {author} {\bibinfo {author} {\bibfnamefont {C.}~\bibnamefont
  {Park}},\ }\href {\doibase 10.1016/j.physletb.2023.137672} {\bibfield
  {journal} {\bibinfo  {journal} {Phys. Lett. B}\ }\textbf {\bibinfo {volume}
  {838}},\ \bibinfo {pages} {137672} (\bibinfo {year} {2023})},\ \Eprint
  {http://arxiv.org/abs/2106.05500} {arXiv:2106.05500 [hep-th]} \BibitemShut
  {NoStop}%
\end{thebibliography}%

\end{document}